\documentclass[journal]{./IEEEtran/IEEEtran}

\usepackage{amsmath} 
\usepackage{amsfonts}
\usepackage{mathtools} 
\usepackage{amssymb}
\usepackage{graphicx}
\usepackage{bm}  % Bold math
\usepackage[colorinlistoftodos, textsize=footnotesize]{todonotes}
\usepackage[colorlinks=true, allcolors=blue]{hyperref}
\usepackage{setspace}
\usepackage{algorithm}      % http://ctan.org/pkg/algorithm
\usepackage{algpseudocode}  % http://ctan.org/pkg/algorithmicx
\usepackage{tikz}
\usepackage{verbatim}
\usepackage{xcolor}
\usepackage{subcaption}

\usepackage{adjustbox}

\usepackage[normalem]{ulem}

\usepackage[thinc]{esdiff}

\usepackage{amsthm}
\usepackage{mathrsfs}

\newtheorem{theorem}{Theorem}

\newtheorem{lemma}{Lemma}
\newtheorem{prop}{Proposition}

\newif\ifcompile
\DeclareMathOperator*{\argmin}{arg\,min}

\newcommand{\reviewquestion}[2]{#2}

\newcommand{\vf}[1]{{\bm{#1}}}
\newcommand{\mf}[1]{{\mathbf{#1}}}
\newcommand{\bgamma}{{\bm\gamma}} 
\newcommand{\bsigma}{{\bm\sigma}}
\newcommand{\bxi}{{\bm\xi}}
\newcommand\defeq{\stackrel{\text{\tiny def}}{=}}

\begin{document}

% Paper Title: Titles are generally capitalized except for words such as a, an,
% and, as, at, but, by, for, in, nor, of, on, or, the, to and up, which are
% usually not capitalized unless they are the first or last word of the title.
% Linebreaks \\ can be used within to get better formatting as desired. Do not
% put math or special symbols in the title.
\title{An Unconstrained Convex Formulation of Compliant Contact}

% Author Names and IEEE memberships: Note positions of commas and nonbreaking
% spaces ( ~ ) LaTeX will not break a structure at a ~ so this keeps an author's
% name from being broken across two lines. Use \thanks{} to gain access to the
% first footnote area a separate \thanks must be used for each paragraph as
% LaTeX2e's \thanks was not built to handle multiple paragraphs.
\author{Alejandro Castro, Frank Permenter, Xuchen Han% <-this % stops a space
\thanks{All the authors are with the Toyota Research Institute, USA, {\tt\small
firstname.lastname@tri.global}.}%
}

%{Shell \MakeLowercase{\textit{et al.}}: Bare Demo of IEEEtran.cls for IEEE
%Journals}

\maketitle

% Abstract:
\begin{abstract}
We present a convex formulation of compliant frictional contact and a robust,
performant method to solve it in practice. By analytically eliminating contact
constraints, we obtain an unconstrained convex problem. Our solver has proven
global convergence and warm-starts effectively, enabling simulation at
interactive rates. We develop compact analytical expressions of contact forces
allowing us to describe our model in clear physical terms and to rigorously
characterize our approximations. Moreover, this enables us not only to model
point contact, but also to incorporate sophisticated models of compliant contact
patches. Our time stepping scheme includes the midpoint rule, which we
demonstrate achieves second order accuracy even with frictional contact. We
introduce a number of accuracy metrics and show our method outperforms existing
commercial and open source alternatives without sacrificing accuracy. Finally,
we demonstrate robust simulation of robotic manipulation tasks at interactive
rates, with accurately resolved stiction and contact transitions, as required
for meaningful sim-to-real transfer. Our method is implemented in the open
source robotics toolkit Drake.

\end{abstract}

% Keywords:

% For peer review papers, you can put extra information on the cover page as
% needed: \ifCLASSOPTIONpeerreview \begin{center} \bfseries EDICS Category:
% 3-BBND \end{center} \fi
%
% For peerreview papers, this IEEEtran command inserts a page break and creates
% the second title. It will be ignored for other modes.
\IEEEpeerreviewmaketitle

% We place each section in its own file:
\section{Introduction}
\label{sec:introduction}

\IEEEPARstart{S}{imulation} of multibody systems with frictional contact has
proven indispensable in robotics, aiding at multiple stages during the
mechanical and control design, testing, and training of robotic systems. Robotic
applications often require robust simulation tools that can perform at
interactive rates without sacrificing accuracy, a critical prerequisite for
meaningful sim-to-real transfer. However, reliable modeling and simulation for
contact-rich robotic applications remains somewhat elusive.

Rigid body dynamics with frictional contact is complicated by the non-smooth
nature of the solutions. It is well known \cite{bib:baraff1993issues} that rigid
contact when combined with the Coulomb model of friction can lead to paradoxical
configurations for which solutions in terms of accelerations and forces do not
exist. These phenomena are known as Painlev\'e paradoxes
\cite{bib:hogan2017regularization}.
\begin{figure}[!ht]
	\centering
    \includegraphics[width=0.95\columnwidth]{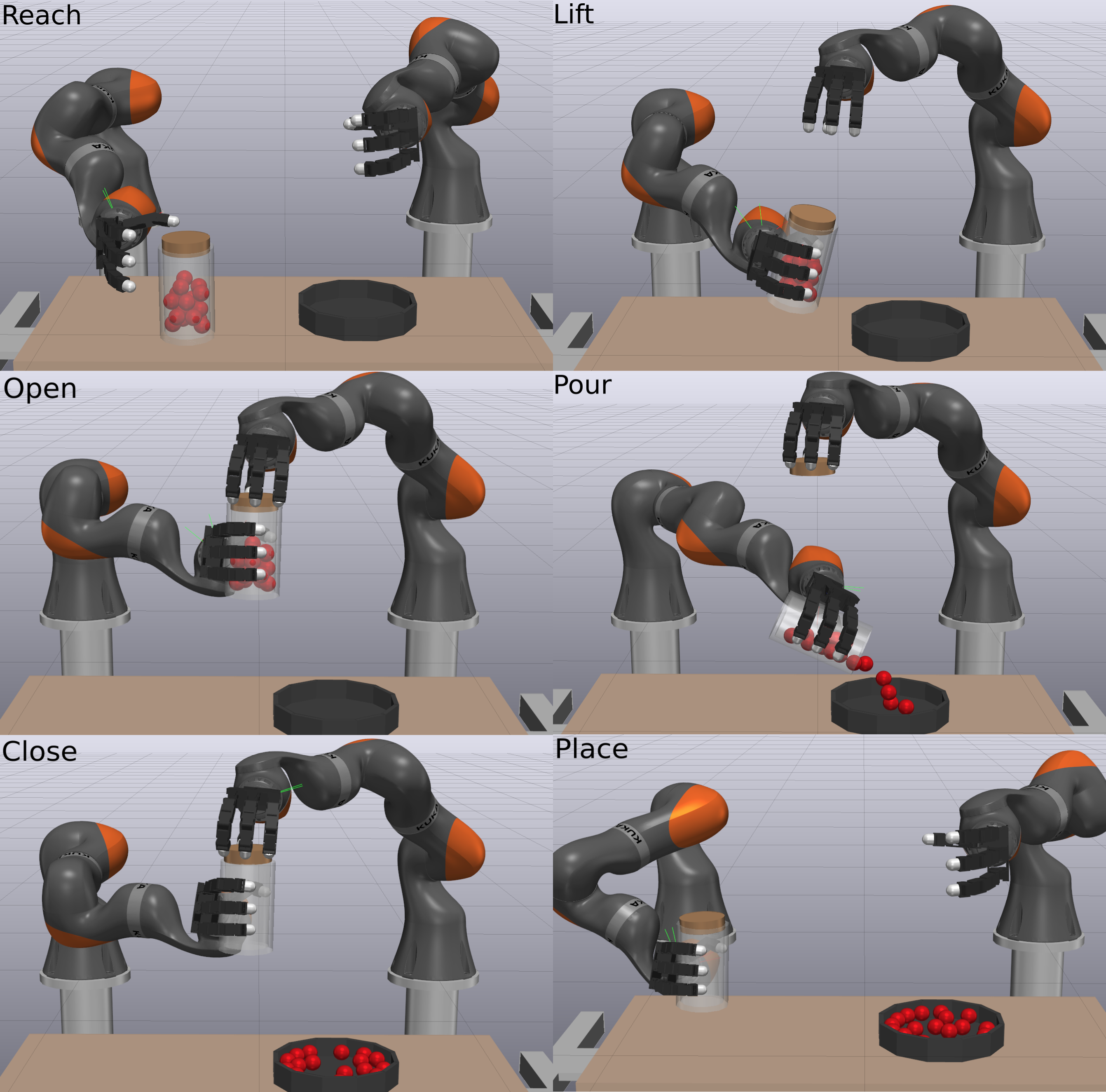}
	\caption{\label{fig:dual_arm_frames}
    Keyframes of a dual arm manipulation task in simulation (see supplemental
    video). The robot is commanded to pick up a jar full of marbles, open it,
    pour its contents into a bowl, close the lid and place the empty jar back in
    place. This is a computationally intensive simulation with 160 degrees of
    freedom and hundreds of contact constraints per time step (see Fig.
    \ref{fig:dual_arm_contacts}). Our SAP solver is robust and warm-starts
    effectively, enabling this simulation to run at interactive rates.}
\end{figure}
Theory resolves these paradoxes by allowing discrete velocity jumps and
impulsive forces, formally casting the problem as a differential variational
inequality \cite{bib:pang2008differential}. In practice, event based approaches
can resolve impulsive transitions \cite{bib:haug1986}, though it is not clear
how to reliably detect these events even for simple one degree of freedom
systems \cite{bib:hogan2017regularization}.

Nevertheless, the problem can be solved in a weaker formulation at the velocity
level using a time-stepping scheme where the next step velocities and impulses
are the unknowns at each time step \cite{bib:stewart1996implicit,
bib:anitescu1997}. These formulations lead in general to a non-linear
complementarity problem (NCP). A linear complementarity problem (LCP) can be
formulated using a polyhedral approximation of the friction cone, though at the
expense of non-physical anisotropy \cite{bib:li2018implicit}. Even though LCP
formulations guarantee solution existence \cite{bib:anitescu1997,
bib:stewart1998convergence}, solving them accurately and efficiently has
remained difficult in practice. This has been explained partly due to the fact
that these formulations are equivalent to nonconvex problems in global
optimization, which are generally NP-hard \cite{bib:Kaufman2008}. Indeed,
popular direct methods based on Lemke's pivoting algorithm to solve LCPs may
exhibit exponential worst-case complexity \cite{bib:baraff1994fast}. Similarly,
popular iterative methods based on projected Gauss-Seidel (PGS)
\cite{bib:duriez2006_realistic_haptic_rendering, bib:bullet} have also shown
exponentially slow convergence \cite{bib:erleben2007velocity}. These
observations are not just of theoretical value---in practice, these methods are
numerically brittle and lack robustness when tasked with computing contact
forces. Software typically attempts to compensate for this inherent lack of
stability and robustness through non-physical constraint relaxation and
stabilization, requiring a significant amount of application-specific parameter
tuning.

\subsection{Previous Work on Convex Approximations of Contact}
To improve computational tractability, Anitescu introduced a \textit{convex
relaxation} of the contact problem \cite{bib:anitescu2006}. This relaxation is a
convex approximation with proven convergence to the solution of a measure
differential inclusion as the time step goes to zero. \reviewquestion{R4-Q1}{For
sliding contacts, the convex approximation introduces a \emph{gliding} artifact
at a distance $\phi$ that is proportional to the time step $\delta t$, friction
coefficient $\mu$ and sliding velocity $\Vert\vf{v}_t\Vert$
\cite{bib:mazhar2014}, i.e. $\phi\sim \delta t\mu\Vert\vf{v}_t\Vert$. This
artifact can be irrelevant for problems with lubricated contacts (as in many
mechanisms) and goes away for sticking contacts (dominant in robotic
manipulation). For sliding contact, the approximation can be adequate for
applications for which the product $\delta t\mu\Vert\vf{v}_t\Vert$ is usually
sufficiently small.} For trajectory optimization, Todorov \cite{bib:todorov2011}
introduces regularization into Anitescu's formulation in order to write a
strictly convex formulation with a unique, smooth and invertible solution. For
simulation, Todorov \cite{bib:todorov2014} uses regularization to introduce
\emph{numerical compliance} that provides Baumgarte-like stabilization to avoid
constraint drift. As a side effect, the regularized formulation can lead to a
noticeable non-zero slip velocity even during stiction \cite{bib:simbenchmark}.

\subsection{Available Software}
Even though these formulations introduce a tractable approximation of frictional
contact, they have not been widely adopted in practice. We believe this is
because of the lack of robust solution methods with a computational cost
suitable for interactive simulation. Software such as ODE \cite{bib:ode}, Dart
\cite{bib:dart} and Vortex \cite{bib:vortex} use a polyhedral approximation of
the friction cone leading to an LCP formulation. Algoryx \cite{bib:algoryx} uses
a \emph{split solver}, reminiscent of one iteration in the staggered projections
method \cite{bib:Kaufman2008}. \reviewquestion{AE-Q3}{RaiSim \cite{bib:raisim}
implements an iterative method with exact solution per-contact, a significant
improvement over the popular PGS used in computer graphics, though still
with no convergence nor accuracy guarantees.} Drake \cite{bib:drake} solves
compliant contact with regularized friction with its transition aware solver
TAMSI \cite{bib:castro2020}.

To our knowledge, Chrono \cite{bib:chrono2016}, Mujoco \cite{bib:mujoco} and
Siconos \cite{bib:acary2019siconos} are the only packages that implement the
convex approximation of contact. \reviewquestion{R4-Q2}{Chrono implements a
variety of solvers for Anitescu's convex formulation \cite{bib:anitescu2006}
including projected Jacobi and Gauss-Seidel methods \cite{bib:tasora2011},
Accelerated Projected Gradient Descent (APGD) \cite{bib:mazhar2015}, Spectral
Projected Gradients (SPG) \cite{bib:heyn2013using} and more recently Alternating
Direction Method of Multipliers (ADMM) \cite{bib:tasora2021solving}. Though
these methods are first order and often exhibit slow convergence, they are
amenable to parallelization and have been applied successfully in the simulation
of granular flows with millions of bodies.} MuJoCo targets robotic applications.
Its contact model is parameterized by a daunting number of non-physical
parameters aimed at regularizing the problem, though at the expense of drift
artifacts \cite{bib:simbenchmark}. Still, it has become very popular in the
reinforcement learning community given its performance.
\reviewquestion{R1-Q8}{Siconos is an open-source software targeting
large scale simulation of both rigid and deformable objects. The authors of
Siconos perform an exhaustive benchmarking campaign in
\cite{bib:acary2018solving} using the convex approximation of contact. The
analysis reveals that there is no universal solver and that every solver
technology suffers from accuracy, robustness and/or efficiency problems.}

\subsection{Outline and Novel Contributions}
\reviewquestion{R1-Q9}{It is not clear if these convex approximations
present a real advantage when compared to approaches solving the original
non-convex NCP problem and whether the artifacts introduced by the approximation
are acceptable for robotics applications.} In this work, we propose a new
unconstrained convex approximation and discuss techniques for its efficient
implementation. We carefully quantify the artifacts introduced by the convex
approximation and evaluate the robustness and accuracy of our method on a family
of examples.

We introduce a two-stage time stepping approach (Section
\ref{sec:discrete_time_formulation}) that allows us to incorporate both first
and second order schemes such as the midpoint rule. In Section \ref{sec:spring_cylinder} we
demonstrate that the midpoint rule can achieve second order accuracy even in
problems with frictional contact. Unlike previous work
\cite{bib:anitescu2010,bib:todorov2014} that formulates the problem in its dual
form (impulses), we write a primal formulation of compliant contact in
velocities (Section \ref{sec:primal_formulation}). We then analytically
eliminate constraints from this formulation to obtain an unconstrained convex
problem (Section \ref{sec:unconstrained_convex_formulation}).

To solve this formulation, we develop SAP---the Semi-Analytic Primal solver---in
Section \ref{sec:sap_solver} and study its theoretical and practical
convergence.  Crucially, we show that SAP globally converges from all initial
conditions (Appendix \ref{app:sap_converge}) and warm-starts effectively using
the previous time-step velocities, enabling simulation at interactive rates.

To address accuracy and model validity,
Section \ref{sec:physical_intuition} provides compact analytic formulae for the
impulses that correspond to the optimal velocities of the convex approximation.
This provides intuition for the approximation, even to those without
optimization expertise. Moreover, the artifacts introduced by the approximation
become apparent and can be characterized precisely. \reviewquestion{AE-Q7}{We
provide an exact mapping between the regularization introduced by Todorov
\cite{bib:todorov2014} to \emph{physical} parameters of compliance. Therefore
regularization is no longer treated as a tuning parameter of the algorithm but
as a true physical parameterization of the contact model. This allows us to
incorporate not only compliant point contact, but also complex models of
compliant surface patches \cite{bib:elandt2019pressure,
bib:masterjohn2021discrete}, as we demonstrate in Section
\ref{sec:slip_control}.}

We demonstrate the effectiveness of our approach in Section \ref{sec:test_cases}
in a number of simulation cases, including the simulation of the challenging
dual arm manipulation task shown in Fig. \ref{fig:dual_arm_frames}. We evaluate
the accuracy, robustness and performance of SAP against available commercial and
open-source optimization solvers.

Finally, we discuss extensions and variations in Section
\ref{sec:variations_and_extensions}, limitations in Section
\ref{sec:limitations} and conclude with final remarks in Section
\ref{sec:future_directions}.

% Dummy comment for Reviewable.

\section{Multibody Dynamics with Contact}
\label{sec:multibody_dynamics_with_contact}

\reviewquestion{R1-Q3}{We use generalized coordinates (in particular joint
coordinates) to describe our multibody system.} Therefore, the state is fully
specified by the generalized positions $\mf{q}\in\mathbb{R}^{n_q}$ and the
generalized velocities $\mf{v}\in\mathbb{R}^{n_v}$, where $n_q$ and $n_v$ denote
the number of generalized positions and velocities, respectively. Time
derivatives of the configurations relate to generalized velocities by the
kinematic map $\mf{N}(\mf{q})\in\mathbb{R}^{n_q\times n_v}$ as
\begin{equation}
    \dot{\mf{q}}=\mf{N}(\mf{q})\mf{v}.
    \label{eq:v_to_qdot}
\end{equation}

% Dummy comment for Reviewable.

\subsection{Contact Kinematics}
\label{sec:contact_modeling}

\reviewquestion{R8-Q1/AE-Q4}{Given a configuration $\mf{q}$ of the system, we
assume our geometry engine reports a set $\mathscr{C}(\mf{q})$ of $n_c$
potential contacts between pairs of bodies. We characterize the $i\text{-th}$
\emph{contact pair} in $\mathscr{C}(\mf{q})$ by the location $\vf{p}_i$ of the
contact point, a normal direction $\hat{\vf{n}}_i$ and \emph{signed distance
function}
\cite{bib:flores2021contact,bib:pfeiffer1996multibody}
$\phi_i(\mf{q})\in\mathbb{R}$, defined negative for overlapping bodies.} The
kinematics of each contact is further completed with the relative velocity
$\vf{v}_{c,i}\in\mathbb{R}^3$ between these two bodies at point $\vf{p}_i$,
expressed in a contact frame $C_i$ for which we arbitrarily choose the
$z\text{-axis}$ to coincide with the contact normal $\hat{\vf{n}}_i$. In this
frame the normal and tangential components of $\vf{v}_{c,i}$ are given by
$v_{n,i} = \hat{\vf{n}}_i\cdot\vf{v}_{c,i}$ and $\vf{v}_{t,i} =
\vf{v}_{c,i}-v_{n,i}\hat{\vf{n}}_i$ respectively, so that
$\vf{v}_{c,i}=[\vf{v}_{t,i}\,v_{n,i}]$.

We form the vector $\mf{v}_{c}\in\mathbb{R}^{3n_c}$ of contact velocities by
stacking velocities $\vf{v}_{c,i}$ of all contact pairs together. In general,
unless otherwise specified, we use bold italics for vectors in $\mathbb{R}^3$
and non-italics bold for their stacked counterparts. The generalized velocities
$\mf{v}$ and contact velocities $\mf{v}_c$ satisfy the equation
$\mf{v}_c=\mf{J}\,\mf{v}$, where $\mf{J}(\mf{q})\in\mathbb{R}^{3n_c\times n_v}$
denotes the contact Jacobian.

% Dummy comment for Reviewable.

\subsection{Contact Modeling}
We model the normal component of the impulse $\gamma_{n}$ during a time
interval of size $\delta t$ with the compliant law
\begin{equation}
    \gamma_{n}/\delta t = (-k\phi - \tau_{d}\,k\,v_{n})_+
    \label{eq:compliant_model}
\end{equation}
where $k$ is the stiffness parameter, $\tau_{d}$ is a \textit{dissipation time
scale} and $(x)_+=\max(0, x)$ is the \textit{positive part} operator.
\reviewquestion{AE-Q5}{The $\delta t$ is needed to convert \emph{forces} into
\emph{impulses}.} This model of compliance can be written as the equivalent
complementarity condition
\begin{equation}
    0 \le \phi + \tau_{d}\,v_{n} + \frac{c}{\delta t} \gamma_{n}\perp \gamma_{n} \ge 0
\end{equation}
where $c=k^{-1}$ is the compliance parameter \reviewquestion{AE-Q6}{and $0 \le
a\perp b \ge 0$ denotes complementarity, i.e. $a \ge 0$, $b \ge 0$ and
$a\,b=0$.}

The tangential component $\bgamma_{t}\in\mathbb{R}^2$ of the contact impulse is
modeled with Coulomb's law of dry friction as
\begin{equation}
    \bgamma_{t}=\argmin_{\Vert\bxi\Vert\leq\mu\gamma_{n}}\vf{v}_{t}\cdot\bxi
    \label{eq:maximum_dissipation_principle}
\end{equation}
where $\mu > 0$ is the coefficient of friction. Equation
(\ref{eq:maximum_dissipation_principle}) describes the \emph{maximum dissipation
principle}, which states that friction impulses maximize the rate of energy
dissipation. In other words, friction impulses oppose the sliding velocity
direction. Moreover, \eqref{eq:maximum_dissipation_principle} states that
contact impulses $\bgamma$ are constrained to be in the friction cone
$\mathcal{F}=\{[\vf{x}_t, x_n] \in\mathbb{R}^3 \,|\, \Vert\vf{x}_t\Vert\le \mu
x_n\}$. The optimality conditions for Eq. (\ref{eq:maximum_dissipation_principle}) are
\cite{bib:stewart2000rigid, bib:tasora2011}
\begin{flalign}
    &\mu\gamma_{n}\vf{v}_{t} + \lambda \bgamma_{t} = \vf{0}\nonumber\\
    &0\le \lambda \perp \mu\gamma_{n}-\Vert\bgamma_{t}\Vert \ge 0
    \label{eq:mpd_optimality_conditions}
\end{flalign}
where $\lambda$ is the multiplier needed to enforce Coulomb's law condition
$\Vert\bgamma_{t}\Vert \le \mu\gamma_{n}$. Notice that in the form we
write Eq. (\ref{eq:mpd_optimality_conditions}), multiplier $\lambda$ is zero
during stiction and takes the value $\lambda=\Vert\vf{v}_{t}\Vert$ during
sliding. Finally, the total contact impulse $\bgamma\in\mathbb{R}^3$ expressed
in the contact frame $C$ is given by $\bgamma=[\bgamma_{t}\,\gamma_{n}]$.

% Dummy comment for Reviewable.

\subsection{Discrete Model}\label{sec:discrete_time_formulation}
We base our time-stepping scheme on the $\theta\text{-method}$ \cite[\S
II.7]{bib:hairer2008solving}. We discretize time into intervals of fixed size
$\delta t$ and seek to advance the state of the system from time $t^n$ to the
next step at $t^{n+1} = t^n + \delta t$. In the $\theta\text{-method}$,
variables are evaluated at intermediate time steps $t^\theta = \theta
t^{n+1}+(1-\theta)t^{n}$, with $\theta \in [0, 1]$. \reviewquestion{R1-Q6}{We define \emph{mid-step
quantities} $\mf{q}^{\theta}$, $\mf{v}^{\theta}$, and
$\mf{v}^{\theta_{vq}}$ in accordance with the standard $\theta\text{-method}$
using scalar parameters $\theta$ and $\theta_{vq}$
\begin{align}
	\mf{q}^{\theta} &\defeq \theta\mf{q} + (1-\theta)\mf{q}_0,\nonumber\\
	\mf{v}^{\theta} &\defeq \theta\mf{v} + (1-\theta)\mf{v}_0,\nonumber\\
	\mf{v}^{\theta_{vq}} &\defeq \theta_{vq}\mf{v} + (1-\theta_{vq})\mf{v}_0,
	\label{eq:theta_method}
\end{align}}
where, to simplify notation, we use the naught subscript to denote quantities
evaluated at the previous time step $t^n$ while no additional subscript is used
for quantities at the next time step $t^{n+1}$. Using these definitions we write
the following time stepping scheme where the unknowns are the next time step
generalized velocities $\mf{v}\in\mathbb{R}^{n_v}$, impulses
$\bgamma\in\mathbb{R}^{3n_c}$ and multipliers ${\bm\lambda}\in\mathbb{R}^{n_c}$
\begin{flalign}
    % Momentum equation.
	&\mf{M}(\mf{q}^{\theta}(\mf{v}))(\mf{v}-\mf{v}_0) =\nonumber\\
	&\qquad\delta
	t\,\mf{k}(\mf{q}^{\theta}(\mf{v}),\mf{v}^{\theta}(\mf{v})) +
	\mf{J}(\mf{q}_0)^T\mf{\bgamma}, \label{eq:scheme_momentum}\\
    % Non-penetration condition.
    &0 \le \phi_i(\mf{q}(\mf{v})) + \tau_{d,i}\,v_{n,i}(\mf{q}, \mf{v}) + \frac{c_i}{\delta t}\gamma_{n,i}\nonumber\\
    &\qquad\perp \gamma_{n,i} \ge 0, \quad\qquad\qquad\qquad i\in\mathscr{C}(\mf{q}_0)
    \label{eq:scheme_nonpenetration}\\
    % Maximum dissipation principle.
    &\mu_i\gamma_{n,i}\vf{v}_{t,i}(\mf{q}, \mf{v}) + \lambda_i \bgamma_{t,i} = \vf{0},
    \!\qquad i\in\mathscr{C}(\mf{q}_0)
    \label{eq:scheme_mdp_multiplier}\\
    &0\le \lambda_i \perp \mu_i\gamma_{n,i}-\Vert\bgamma_{t,i}\Vert \ge 0
    , \!\!\qquad i\in\mathscr{C}(\mf{q}_0)
    \label{eq:scheme_mdp_cone}\\
    % Positions update.
    &\dot{\mf{q}}^{\theta_{vq}}(\mf{v}) = \mf{N}(\mf{q}^{\theta}(\mf{v}))\mf{v}^{\theta_{vq}}(\mf{v}),\label{eq:qdot_map_to_v}\\    
    &\mf{q}(\mf{v}) = \mf{q}_0 + \delta t \dot{\mf{q}}^{\theta_{vq}}(\mf{v}),
    \label{eq:scheme_q_update}
\end{flalign}
where $\mf{M}(\mf{q})\in\mathbb{R}^{n_v\times n_v}$ is the mass matrix and
$\mf{k}(\mf{q},\mf{v})\in\mathbb{R}^{n_v}$ models external forces such as
gravity, gyroscopic terms and other smooth generalized forces such as those
arising from springs and dampers.

\reviewquestion{R1-Q6}{This scheme includes some of the most popular schemes for
forward dynamics:
\begin{itemize}
	\item Explicit Euler with $\theta=\theta_{vq} = 0$,
	\item Symplectic Euler with $\theta = 0$ and $\theta_{vq}=1$,
	\item Implicit Euler with $\theta = \theta_{vq}= 1$, and
	\item Symplectic midpoint rule, which is second order, with $\theta =
\theta_{vq}= 1/2$.
\end{itemize}

Notice that in our version of the $\theta\text{-method}$, the additional
parameter $\theta_{vq}$ allows us to also incorporate the popular Symplectic
Euler scheme}.

When only conservative forces are considered in
$\mf{k}(\mf{q},\mf{v})$, the symplectic Euler scheme keeps the total mechanical
energy bounded while exact energy conservations can be attained with the second
order midpoint rule, see results in Section \ref{sec:spring_cylinder}. In
addition, stability analysis in \cite{bib:anitescu2002,bib:potra2006linearly}
shows that these implicit schemes are appropriate for the integration of stiff
forces arising in multibody applications such as springs and dampers.

% Dummy comment for Reviewable.

\subsection{Two-Stage Scheme}
\label{sec:two_stage_scheme}

Similar to the work in \cite{bib:duriez2005realistic} for the simulation of
deformable objects and to projection methods used in fluid mechanics
\cite{bib::bell1991efficient}, we solve Eqs.
(\ref{eq:scheme_momentum})-(\ref{eq:scheme_q_update}) in two stages. In the
first stage, we solve for the \emph{free motion velocities} $\mf{v}^*$ the
system would have in the absence of contact constraints, according to
\begin{align}
	\mf{m}(\mf{v}^*) &= \mf{0},
	\label{eq:vstar_definition}
\end{align}
where we define the momentum residual $\mf{m}(\mf{v})$ from Eq.
(\ref{eq:scheme_momentum}) as
\begin{equation}
	\mf{m}(\mf{v}) =
	\mf{M}(\mf{q}^{\theta}(\mf{v}))(\mf{v}-\mf{v}_0) -
	\delta t\,\mf{k}(\mf{q}^{\theta}(\mf{v}), \mf{v}^{\theta}(\mf{v})).
	\label{eq:m_definition}
\end{equation}

For integration schemes that are implicit in $\mf{v}^*$ (e.g. the implicit Euler
scheme and the midpoint rule), we solve Eq. (\ref{eq:vstar_definition}) with
Newton's method. For schemes explicit in $\mf{v}^*$, only the mass matrix
$\mf{M}$ needs to be inverted, which can be accomplished efficiently using the
$\mathcal{O}(n)$ \emph{Articulated Body Algorithm}
\cite{bib:featherstone2008_rigid_body_dynamics_algorithms}.

The second stage solves a linear approximation of the balance of momentum in Eq.
(\ref{eq:scheme_momentum}) about $\mf{v}^*$ that satisfies the contact
constraints, Eqs. (\ref{eq:scheme_nonpenetration})-(\ref{eq:scheme_mdp_cone}). To write
a convex formulation of contact in Section \ref{sec:convex_approximation}, our
linearization uses a symmetric positive definite (SPD) approximation of
the Jacobian $\partial \mf{m}/\partial \mf{v}$. To achieve this, we split the
non-contact forces $\mf{k}$ in Eq. (\ref{eq:scheme_momentum}) as
\begin{equation*}
	\mf{k}(\mf{q}, \mf{v}) = \mf{k}_1(\mf{q}, \mf{v})+\mf{k}_2(\mf{q}, \mf{v}),
\end{equation*}
such that the Jacobians $\partial \mf{k}_1/\partial\mf{q}$ and $\partial
\mf{k}_1/\partial\mf{v}$ are negative definite matrices while the same is
generally not true for the Jacobians of $\mf{k}_2$. The term $\mf{k}_1(\mf{q},
\mf{v})$ can include forces from modeling elements such as spring and dampers.
The term $\mf{k}_2(\mf{q}, \mf{v})$ includes all other contributions that cannot
guarantee negative definiteness of their Jacobians, such as Coriolis and
gyroscopic forces arising in multibody dynamics with generalized coordinates. We
can now define the SPD approximation of $\partial \mf{m}/\partial \mf{v}$
evaluated at $\mf{v}^*$ as
\begin{align}
	\mf{A}&=\mf{M}+\delta t^2\,\theta\theta_{qv}\mf{K}+\delta t\,\theta\mf{D},
	\label{eq:expression_for_A}\\
	\mf{K}&=-\frac{\partial \mf{k}_1}{\partial
	\mf{q}}\frac{\partial\dot{\mf{q}}^{\theta_{vq}}}{\partial\mf{v}},
	\label{eq:stiffness_matrix}\\
	\mf{D}&=-\frac{\partial \mf{k}_1}{\partial\mf{v}},
	\label{eq:dissipation_matrix}
\end{align}
\reviewquestion{R4-Q3}{where $\mf{K} \succ 0$ and $\mf{D}\succ 0$ are the stiffness and damping
matrices of the system, respectively. For joint level
spring-dampers models, $\mf{K}$ and $\mf{D}$ are constant, diagonal, and
positive definite matrices. Section \ref{sec:spring_cylinder} shows the
performance of our scheme for a system with a linear spring.}

Using this SPD approximation of $\partial \mf{m}/\partial \mf{v}$, the
linearized balance of momentum \eqref{eq:scheme_momentum} reads
\begin{equation}
    % Momentum equation.
	\mf{A}(\mf{v}-\mf{v}^*) = \mf{J}^T\mf{\bgamma},
	\label{eq:momentum_linearized}
\end{equation}
where for convenience we use $\mf{J}$ as a shorthand to denote
$\mf{J}(\mf{q}_0)$. The approximation in Eq. (\ref{eq:momentum_linearized}) and
the original discrete momentum update in Eq. (\ref{eq:scheme_momentum}) agree to
second order as shown by the following result, proved in Appendix
\ref{app:gradient_of_m_approximation}.
\begin{prop}	
Matrix $\mf{A}$ is a first order approximation to the Jacobian of $\mf{m}$,
i.e.,
\begin{align*}
	\left. \frac{\partial \mf{m}}{\partial \mf{v}} \right|_{\mf{v}=\mf{v}^*} = \mf{A} + \mathcal{O}(\delta t).
\end{align*}
Therefore, Eq. (\ref{eq:momentum_linearized}) is a second order approximation of
the discrete balance of momentum in Eq. (\ref{eq:scheme_momentum}). Moreover,
$\mf{A} \succ 0$.
\label{prop:gradient_of_m_approximation}
\end{prop}

Notice that, in the absence of constraint impulses, the velocities at the next
time step are equal to the free motion velocities, i.e., $\mf{v}=\mf{v}^*$, and
they are computed with the order of accuracy of the $\theta\text{-method}$.
Furthermore, we also expect to recover the properties of the
$\theta\text{-method}$ when contact constraints are in stiction. As an example,
for bodies in contact that are under rolling friction, the contact constraints
behave as bi-lateral constraints that impose zero slip velocity. In this case,
our two-stage method with the midpoint rule exhibits considerably less numerical
dissipation than first order methods (see results in Section
\ref{sec:spring_cylinder}.)

Even after the linearization of the balance of momentum in Eq.
\eqref{eq:momentum_linearized}, the full problem with the contact constraints
\eqref{eq:scheme_nonpenetration}-\eqref{eq:scheme_mdp_cone} still consists of a
non-convex nonlinear complementarity problem (NCP). This kind of problems are
difficult to solve in practice, especially so in engineering applications for
which robustness and accuracy are required. In the next section we introduce a
number of approximations to this original NCP that allow us to make the problem
tractable and solve it efficiently in practice.

% Dummy comment for Reviewable.

\section{Convex Approximation of Contact Dynamics}
\label{sec:convex_approximation}

In this section we build from previous work on convex approximations of contact
\cite{bib:anitescu2006, bib:todorov2011, bib:todorov2014} to write a new convex
formulation of \emph{compliant} contact in terms of \emph{velocities}.

% Dummy comment for Reviewable.

\subsection{Primal Formulation}
\label{sec:primal_formulation}

We introduce a new decision variable $\vf{\sigma}\in\mathbb{R}^{3n_c}$ and set
up our convex formulation of compliant contact as the following 
optimization problem
\begin{equation}
	\begin{aligned}
	\min_{\mf{v},\bsigma} \quad & \ell_p(\mf{v},\bsigma) =
	\frac{1}{2}\Vert\mf{v}-\mf{v}^*\Vert_{A}^2 +
	\frac{1}{2} \Vert\bsigma\Vert_{R}^2\\
	\textrm{s.t.} \quad & \mf{g} = (\mf{J}\mf{v}-\hat{\mf{v}}_c + \mf{R}\bsigma) \in \mathcal{F}^*,\\
	\end{aligned}
	\label{eq:primal_regularized}
\end{equation}
where $\Vert\mf{z}\Vert_X^2=\mf{z}^T\mf{X}\mf{z}$ with $\mf{X}\succ 0$ and
$\mathcal{F^*}= \mathcal{F}^*_1 \times \mathcal{F}^*_2 \times \cdots \times
\mathcal{F}^*_{n_c}$ is the \emph{dual cone} of the friction cone $\mathcal{F} =
\mathcal{F}_1 \times \mathcal{F}_2 \times \cdots \times \mathcal{F}_{n_c}$, with
$\times$ the Cartesian product. The positive diagonal matrix
$\mf{R}\in\mathbb{R}^{3n_c\times 3n_c}$ and the vector of stabilization
velocities $\hat{\mf{v}}_c$ encode the problem data needed to model compliant
contact. We establish a very clear physical meaning for these terms in
Section~\ref{sec:physical_intuition}.  We note that when $R$ and $\bsigma$
are removed, this reformulation reduces to~\cite{bib:mazhar2014}.

We refer to~\eqref{eq:primal_regularized} as the \emph{primal formulation}
and next derive its dual. To begin, define the Lagrangian
\begin{equation}
    \mathcal{L}(\mf{v},\bsigma,\vf{\gamma}) = 
\frac{1}{2}\Vert\mf{v}-\mf{v}^*\Vert_A^2 + \frac{1}{2} \Vert\bsigma\Vert_{R}^2 - \vf{\gamma}^T\mf{g},
\end{equation}
where $\vf{\gamma}\in\mathcal{F}$ is the dual variable for the constraint
$\vf{g}\in \mathcal{F}^*$. Taking gradients of the Lagrangian with respect to $\mf{v}$ and $\bsigma$ leads
to the optimality conditions
\begin{subequations}\label{eq:primal_optimality_conditions}
\begin{align}
    \mf{A}(\mf{v}-\mf{v}^*) &= \mf{J}^T\vf{\gamma} \label{eq:momentum_optimality}\\
    \vf{\sigma} &= \vf{\gamma}.  \label{eq:sigma_equal_gamma}
\end{align}
\end{subequations}
Note that~\eqref{eq:momentum_optimality} is a restatement
of the balance of momentum \eqref{eq:momentum_linearized}
if the dual variable $\bgamma$ is interpreted as a vector of contact impulses. 
 Using the optimality
condition~\eqref{eq:sigma_equal_gamma} to eliminate $\vf{\sigma}$ from the
Lagrangian yields the dual formulation
\begin{align}
    \min_{\bgamma\in \mathcal{F}} \ell_d(\bgamma) =
    \frac{1}{2}\bgamma^T(\mathbf{W}+\mathbf{R})\bgamma + {\bm r}^T
    \bgamma,
	\label{eq:dual_regularized}
\end{align}
where $\mf{W}=\mf{J}\mf{A}^{-1}\mf{J}^T$ is the Delassus operator and
$\mf{r}=\mf{v}_c^*-\hat{\mf{v}}_c$ with $\mf{v}_c^*=\mf{J}\mf{v}^*$.
In summary, we have proven the following theorem.
\begin{theorem}\label{th:primal_dual} The dual of~\eqref{eq:primal_regularized}
	is given by~\eqref{eq:dual_regularized}. Moreover, when $\{\mf{v},\bsigma\}$ is
	primal optimal and $\bgamma$ is dual optimal, $\bsigma = \bgamma$.
\end{theorem}

Finally, we note that the dual~\eqref{eq:dual_regularized} is equivalent to the
formulation in~\cite{bib:todorov2011} when $\theta = 0$ in
Eq.~\eqref{eq:scheme_momentum} and $\mf{A}=\mf{M}(\mf{q}_0)$. 

% Dummy comment for Reviewable.

\subsection{Analytical Inverse Dynamics}
\label{sec:analytical_inverse_dynamics}

The dual optimal impulses of~\eqref{eq:dual_regularized} can be
constructed from the primal optimal velocities of~\eqref{eq:primal_regularized}
using a simple projection operation. Following~\cite{bib:todorov2014}, we call
this construction  \textit{analytical inverse dynamics}. Moreover, this
projection decomposes into a set of individual projections for each contact
impulse $\bgamma_i$ given the separable structure of the constraints. Letting
$\vf{y}_i(\vf{v}_{c,i}) = -\vf{R}_i^{-1}(\vf{v}_{c,i}-\hat{\vf{v}}_{c,i})$,
these projections take the form
\begin{equation}
  \begin{aligned}
	\bgamma_i(\vf{v}_{c,i})&= P_{\mathcal{F}_i}(\vf{y}_i(\vf{v}_{c,i}))\\
	&= \argmin_{\bgamma\in\mathcal{F}_i} \quad 
		\frac{1}{2}(\bgamma-\vf{y}_i)^T\vf{R}_i(\bgamma-\vf{y}_i),\\
	\end{aligned}
	\label{eq:y_projection}
\end{equation}
where $\vf{R}_i\in\mathbb{R}^{3\times3}$ is the $i\text{-th}$ diagonal block of
the regularization matrix $\mf{R}$. That is, $\bgamma_i$ is the projection
$P_{\mathcal{F}_i}$ of $\vf{y}_i(\vf{v}_{c,i})$ onto the friction cone
$\mathcal{F}_i$ using the norm defined by $\vf{R}_i$.
\reviewquestion{R1-Q7}{Remarkably, the projection map $P_{\mathcal{F}_i}$ can be
evaluated \emph{analytically}. We provide algebraic expressions for it in
Section~\ref{sec:physical_intuition} and derivations in
Appendix~\ref{app:analytical_inverse_dynamics_derivations}.} The projection
$P_{\mathcal{F}}(\mf{y})$ onto the full cone $\mathcal{F} := \mathcal{F}_1
\times \mathcal{F}_2 \times \cdots \times \mathcal{F}_{n_c}$ is obtained by
simply stacking together the individual projections
$P_{\mathcal{F}_i}(\vf{y}_i)$ from Eq. (\ref{eq:y_projection}), where we form
$\mf{y}$ by stacking together each $\vf{y}_i$ from all contact pairs.  In this
notation, the optimal impulse $\bgamma$ of~\eqref{eq:dual_regularized} and the
optimal velocities $\mf{v}$ of~\eqref{eq:primal_regularized} satisfy  $\bgamma =
P_{\mathcal{F}}(\mf{y}(\mf{v}))$.

% Dummy comment for Reviewable.

\algblockdefx{RepeatUntil}{EndRepeatUntil}{\textbf{repeat until}}{}
\algnotext{EndRepeatUntil}

\subsection{An Unconstrained Convex Formulation}
\label{sec:unconstrained_convex_formulation}

\reviewquestion{R1-Q7}{We use the analytical $\bgamma =
P_{\mathcal{F}}(\mf{y}(\mf{v}))$ from Section
\ref{sec:analytical_inverse_dynamics} and the optimality condition $\vf{\sigma}
= \bgamma$ from Theorem~\ref{th:primal_dual} to eliminate both $\vf{\sigma}$ and
the constraints from the primal formulation \eqref{eq:primal_regularized}. In
total, we obtain the following unconstrained problem in velocities only}
\begin{eqnarray}
	\min_{\mf{v}} \ell_p(\mf{v}) = \frac{1}{2}\Vert\mf{v}-\mf{v}^*\Vert_{A}^2 +
	\frac{1}{2}\Vert P_\mathcal{F}(\mf{y}(\mf{v}))\Vert_R^2.
	\label{eq:primal_unconstrained}
\end{eqnarray}
Correctness of this reformulation is asserted by the following theorem, which we
prove in Appendix \ref{app:unconstrained_formulation_equivalance}.
\begin{theorem}
  If  $\mf{v}$ solves the unconstrained formulation
    (\ref{eq:primal_unconstrained}), then $(\mf{v}, \bsigma)$, with $\bsigma
    =P_\mathcal{F}(\mf{y}(\mf{v}))$, solves the primal formulation
    (\ref{eq:primal_regularized}).
    \label{th:unconstrained_formulation_equivalance}
\end{theorem}
Lemma \ref{lem:PropertiesOfObj} in Appendix \ref{app:sap_converge} shows that
the unconstrained cost $\ell_p(\mf{v})$ is strongly convex and differentiable
with Lipschitz continuous gradients. Therefore, the unconstrained formulation
(\ref{eq:primal_unconstrained}) has a unique solution, and can be efficiently
solved.  
Section \ref{sec:sap_solver} presents our novel SAP solver specifically designed
for its solution.

We outline our time-stepping scheme in Algorithm \ref{alg:sap_time_stepping}.
\begin{algorithm}
	\caption{Overall Time-Stepping Strategy}
	  \label{alg:sap_time_stepping}
	  \begin{algorithmic}[1]
		  \State Solve free motion velocities from
		  $\mf{m}(\mf{v}^*)=\mf{0}$, Eq. \eqref{eq:m_definition}
		  \State Solve $\displaystyle \mf{v} = \argmin_{\mf{v}} \ell_p(\mf{v})$, Eq. \eqref{eq:primal_unconstrained}
		  \State Update positions $\displaystyle \mf{q} = \mf{q}_0 + \delta
		  t\mf{N}(\mf{q}^{\theta})\mf{v}^{\theta_{vq}}$, Eqs. \eqref{eq:qdot_map_to_v}-\eqref{eq:scheme_q_update}
	  \end{algorithmic}
\end{algorithm}

\reviewquestion{R1-Q1/R1-Q2}{We note that Algorithm \ref{alg:sap_time_stepping}
is executed once per time step, with no inner iterations updating $\mf{A}$. That
is, our strategy is not solving the original, possibly nonlinear, balance of
momentum \eqref{eq:scheme_momentum} but its (second-order accurate) linear
approximation in \eqref{eq:momentum_linearized}. This approximation can be exact
for many multibody systems encountered in practice. For instance, joint springs
and dampers contribute constant stiffness and damping matrices in
\eqref{eq:expression_for_A}.}

\algblockdefx{RepeatUntil}{EndRepeatUntil}{\textbf{repeat until}}{}
\algnotext{EndRepeatUntil}

\section{Semi-Analytic Primal Solver}
\label{sec:sap_solver}

Inspired by Newton's method, \reviewquestion{R1-Q4}{our Semi-Analytic Primal
Solver (SAP) seeks to solve \eqref{eq:primal_unconstrained} by 
monotonically decreasing the primal cost $\ell_p(\mf{v})$} at each iteration, as
outlined in Algorithm \ref{alg:sap}.
\begin{algorithm}[H]
	\caption{The Semi-Analytic Primal Solver (SAP)}	
	  \label{alg:sap}
	  \begin{algorithmic}[1]
		  \State Initialize $\mf{v}_m \gets \mf{v}_0$ \RepeatUntil
		  $~\Vert\tilde{\nabla}\ell_p\Vert < \varepsilon_a +
		  \varepsilon_r\max(\Vert\tilde{\mf{p}}\Vert,\Vert\tilde{\mf{j}_c}\Vert)$,
		  Eq. \eqref{eq:stopping_criteria} \State $\Delta\mf{v}_{m} =
		  -\mf{H}^{-1}(\mf{v}_m)\nabla_\mf{v}\ell_p(\mf{v}_m)$
		  \label{op:Newton_iteration} \State $\displaystyle \alpha_m =
		  \argmin_{t\in\mathbb{R}^{++}} \ell_p(\mf{v}_m + t \Delta\mf{v}_{m})$
		  \State $\displaystyle \mf{v}_{m+1} = \vf{v}_m +
		  \alpha_{m}\Delta\mf{v}_{m}$ \EndRepeatUntil \State\Return $\{\mf{v}$,
		  $\bgamma=P_\mathcal{F}(\vf{y}(\mf{v}))\}$
	  \end{algorithmic}
\end{algorithm}

The SAP iterations require $\mf{H}(\mf v)\succ 0$ at all iterations.  At points
where $\nabla_\mf{v}\ell_p(\mf{v})$ is differentiable, $\mf{H}(\mf v)$ is simply
set equal to the Hessian of the cost function. In general, $\mf{H}(\mf v)$ is
evaluated using a partition of its domain. For each set in the partition,
$\nabla_\mf{v}\ell_p(\mf{v})$ is differentiable on the interior, and the Hessian
admits a simple formula. We globally define $\mf{H}(\mf v)$ by adopting one of
these Hessian formulas on the boundary. The partition is described in Appendix
\ref{app:analytical_inverse_dynamics_derivations}. The Hessian formula are given
in Appendix \ref{app:gradients_derivation}. Our convergence analysis in Appendix
\ref{app:sap_converge} also accounts for this definition.

\reviewquestion{R1-Q4}{As shown in Appendix \ref{app:sap_converge}, SAP
globally converges at least at a linear-rate. Further, SAP exhibits quadratic
convergence when $\nabla^2 \ell_p$ exists in a neighborhood of the optimal
$\mf{v}$. In practice, we initialize SAP with the previous time-step velocity
$\mf{v}_0$.} The stopping criteria is discussed below in Section
\ref{sec:stopping_criteria}.

\subsection{Gradients}
\label{sec:gradients}

We provide a detailed derivation of the gradients in Appendix
\ref{app:gradients_derivation}. Here we summarize the main results required for
implementation. The gradient of the primal cost $\ell_p$ reduces to the balance
of momentum
\begin{equation*}
	\nabla_\mf{v}\ell_p(\mf{v}) = \mf{A}(\mf{v}-\mf{v}^*) - \mf{J}^T\bgamma(\mf{v}),
\end{equation*}
where $\bgamma(\mf{v})=P_\mathcal{F}(\vf{y}(\mf{v}))$ is given by the analytical
inverse dynamics \eqref{eq:analytical_y_projection}. We define matrix
$\mf{G}\succeq 0$ that evaluates to $-\nabla_{\mf{v}_c}\bgamma$ where
$P_\mathcal{F}(\mf{y}(\mf{v}))$ is differentiable. Otherwise $\mf{G}$ extends
our analytical expressions as outlined in Appendix
\ref{app:gradients_derivation}. Matrix $\mf{G}$ is a block diagonal matrix where
each diagonal block for the $i\text{-th}$ contact is a $3\times 3$ matrix.

In total, we evaluate $\mf{H}$ via
\begin{equation*}
	\mf{H} = \mf{A} + \mf{J}^T\mf{G}\,\mf{J},
\end{equation*}
which is strictly positive definite since $\mf{A}\succ 0$.

\subsection{Line Search}

The line search algorithm is critical to the success of SAP given that $\nabla
\ell_p(\mf{v})$ can rapidly change during contact-mode transitions.  We explore
two line search algorithms: an approximate backtracking line search with
Armijo's stopping criteria and an exact (to machine epsilon) line search.

At the $m\text{-th}$ Newton iteration, backtracking line search starts with a
maximum step length of $\alpha_\text{Max}$ and progressively decreases it by a
factor $\rho \in (0, 1)$ as $\alpha\gets\rho\alpha$ until Armijo's criteria
\cite[\S 3.1]{bib:nocedal2006numerical} is satisfied. We write Armijo's criteria
as $~\ell_p(\mf{v}^m + \alpha \Delta\mf{v}^{m}) < \ell_p(\mf{v}^m) +
c\,\alpha\,d\ell_p/d\alpha(\mf{v}^m)$. Typical parameters we use are $\rho=0.8$,
$c=10^{-4}$ and $\alpha_\text{Max}=1.25$.

For the exact line search we use the method \verb;rtsafe; \cite[\S
9.4]{bib:numerical_recipes} to find the unique root of $d\ell/d\alpha$. This is
a one-dimensional root finder that uses the Newton-Raphson method and switches
to bisection when an iterate falls outside a search bracket or when convergence
is slow. The fast computation of derivatives we show next allows us to iterate
$\alpha$ to machine precision at a negligible impact on the computational cost.
In practice, this is our preferred algorithm since it allows us to use very low
regularization parameters without having to tune tolerances in the line search.

\subsection{Efficient Analytical Derivatives For Line Search}

The algorithm \verb;rtsafe; requires the first and second directional
derivatives of $\ell_p$. We show how to compute these
derivatives efficiently in $\mathcal{O}(n)$ operations. Defining $\ell(\alpha) =
\ell_p(\mf{v}+\alpha\Delta\mf{v})$, we compute first and second derivatives as
\begin{align*}
	\frac{d\ell}{d\alpha}&=\Delta\mf{v}^T\nabla_\mf{v}\ell(\alpha),\\
	\frac{d^2\ell}{d\alpha^2}&=\Delta\mf{v}^T\mf{H}(\alpha)\Delta\mf{v}.
\end{align*}

Using the gradients from Section \ref{sec:gradients}, we can write
\begin{align*}
	\frac{d\ell}{d\alpha}(\alpha)=\Delta\mf{v}^T\mf{A}(\mf{v}(\alpha)-\mf{v}^*)-\Delta\mf{v}^T\mf{J}^T\bgamma.
\end{align*}
These are computed efficiently by first calculating the change in velocity
$\Delta\mf{v}_c:=\mf{J}\Delta\mf{v}$ and change of momentum $\Delta\mf{p} :=
\mf{A}\Delta\mf{v}$. The calculation is then completed via 
\begin{align*}
	\frac{d\ell}{d\alpha}(\alpha)=\Delta\mf{p}^T(\mf{v}(\alpha)-\mf{v}^*)
	-\Delta\mf{v}_c^T\bgamma(\alpha),
\end{align*}
which only requires dot products that can be computed in $\mathcal{O}(n_v)$ and
$\mathcal{O}(n_c)$ respectively. Similarly for the second derivatives
\begin{align*}
	\frac{d^2\ell}{d\alpha^2}(\alpha)=\Delta\mf{v}^T\Delta\mf{p} + \Delta\mf{v}_c^T
	\mf{G}(\alpha)\Delta\mf{v}_c.
\end{align*}

Notice the first term on the right can be precomputed before the line search
 starts, while the second term only involves $\mathcal{O}(n_c)$ operations given
 the block diagonal structure of $\mf{G}$.

\subsection{Problem Sparsity}
\label{sec:problem_sparsity}

\begin{figure}[!h]
	\centering
	\includegraphics[width=0.9\columnwidth]{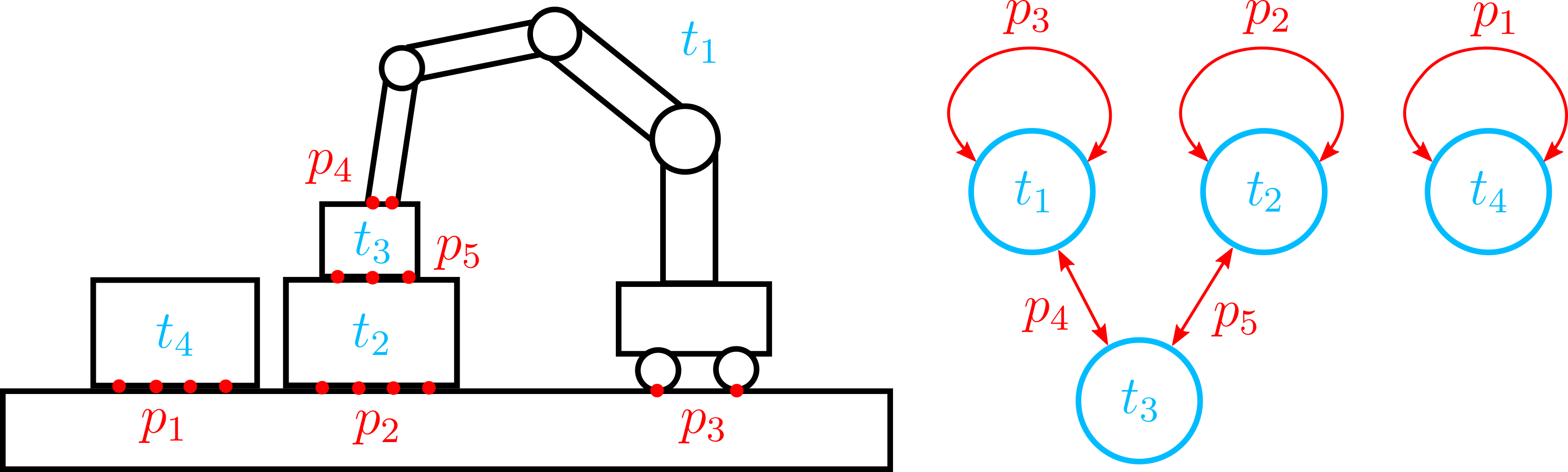}
	\caption{\label{fig:sparsity_example} 
	An example of a sparsity pattern commonly encountered in the simulation of
	robotic mechanical systems. The graph on the right puts \textit{trees} as
	nodes and contact \textit{patches} as edges.}
\end{figure}

The block sparsity of $\mf{H}$  is best described with an example. We organize
our multibody systems as a collection of articulated \emph{tree structures}, or
a \emph{forest}. Consider the system in Fig. \ref{fig:sparsity_example}.
In this example, a robot arm mounted on a mobile base constitutes its own tree,
here labeled $t_1$. The number of degrees of freedom of the $t\text{-th}$ tree
will be denoted with $n_t$. A free body is a common case of a tree with
$n_t=6$. In general, matrix $\mf{A}$ has a block diagonal structure where each
diagonal block corresponds to a tree.

We define as \textit{patches} a collection of contact pairs between two trees.
Each contact pair corresponds to a single cone constraint in our formulation.
The set of constraint indexes that belong to patch $p$ is denoted with
$\mathcal{I}_p$ of size (cardinality) $|\mathcal{I}_p| = r_p$.
Figure~\ref{fig:sparsity_example} shows the corresponding graph where nodes
correspond to trees and edges correspond to patches.

\begin{figure*}[!h]
	\centering
	\includegraphics[width=0.9\textwidth]{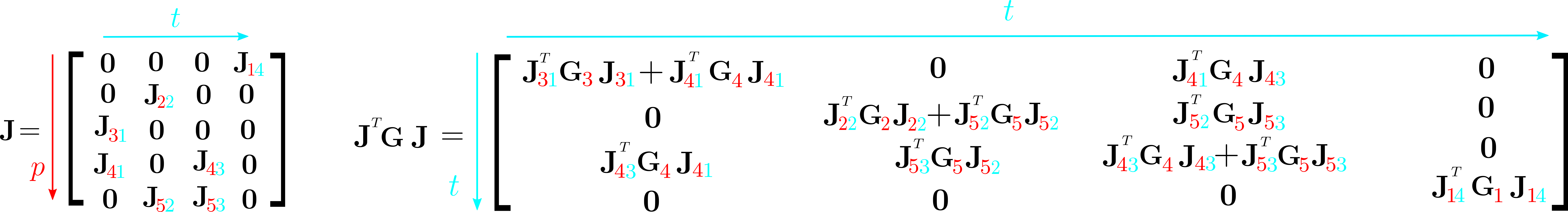}
	\caption{\label{fig:JTGJ_schematic} 
	Block sparsity of the contact Jacobian $\mf{J}$ and the Hessian term
	$\mf{J}^T\mf{G}\mf{J}$, for the example illustrated in Fig.
	\ref{fig:sparsity_example}.}
\end{figure*}

Generally, the Jacobian is sparse since the relative contact velocity only
involves velocities of two trees in contact, Fig. \ref{fig:JTGJ_schematic}. Each
non-zero block $\mf{J}_{pt}$ has size $3r_p\times n_t$. Since $\mf{A}$ is block
diagonal, $\mf{H}$ inherits the sparsity structure of $\mf{J}^T\mf{G}\mf{J}$.

We exploit this structure using a supernodal Cholesky factorization \cite[\S
9]{bib:davis2016survey} that can take advantage of dense algebra optimizations.
Implementing this factorization requires construction of a \emph{junction tree}.
For this we apply the algorithm in \cite{bib:smail2017junction}, using cliques
of $\mf{H}$ as input. We use the implementation from the Conex solver
\cite{bib:permenter2020}.

\reviewquestion{R4-Q7}{The scalability of SAP, like any second-order
optimization method, depends on the complexity of solving the Newton system. For
dense problems, this has $\mathcal{O}(n_v^3)$ complexity, where $n_v$ denotes the
number of generalized velocities. For sparse problems the complexity can be
dramatically reduced \cite{bib:davis2016survey}.  We study scalability with
number of bodies in Section \ref{sec:clutter}. Recent work on the modeling of
contact rich patches \cite{bib:masterjohn2021discrete} studies the scalability
of SAP with the number of constraints.}

\subsection{Stopping Criteria}
\label{sec:stopping_criteria}

To assess convergence, we monitor the norm of the optimality condition for the
unconstrained problem (\ref{eq:primal_unconstrained})
\begin{equation*}
	\nabla\ell_p(\mf{v}) = \mf{A}(\mf{v}-\mf{v}^*) - \mf{J}^T\bgamma.
\end{equation*}

Notice that the components of $\nabla\ell_p$ have units of generalized momentum
$\mf{p}=\mf{M}\mf{v}$. Depending on the choice of generalized coordinates, the
generalized momentum components may have different units. In order to weigh all
components equally, we define the diagonal matrix $\mf{D} =
\text{diag}(\mf{M})^{-1/2}$ and perform the following change of variables
\begin{align}
	\tilde{\nabla}\ell_p &= \mf{D}\nabla\ell_p, \nonumber\\
	\tilde{\mf{p}} &= \mf{D}\mf{p}, \nonumber \\
	\tilde{\mf{j}}_c &= \mf{D}\mf{j}_c,
	\label{eq:scaled_momentum_quantities}
\end{align}
where we define the generalized contact impulse $\mf{j}_c=\mf{J}^T\bgamma$.
With this scaling, all the new \emph{tilde} variables have the same units,
square root of Joules. Using these definitions, we write our stopping criteria
as
\begin{equation}
	\Vert\tilde{\nabla}\ell_p\Vert < \varepsilon_a + \varepsilon_r\max(\Vert\tilde{\mf{p}}\Vert,\Vert\tilde{\mf{j}_c}\Vert).
	\label{eq:stopping_criteria}
\end{equation}
where $\varepsilon_r$ is a dimensionless relative tolerance that we usually set
in the range from $10^{-6}$ to $10^{-1}$. The absolute tolerance $\varepsilon_a$
is used to detect rare cases where the solution leads to no contact and no
motion, typically due to external forces. We always set this tolerance to a
small number, $\varepsilon_a=10^{-16}$.

% Dummy comment for Reviewable.

\section{Contact Modeling Parameters}
\label{sec:contact_modeling_parameters}

Thus far, $\vf{R}_i$ and $\hat{\vf{v}}_{c,i}$ have been treated as known problem
data. This section makes an explicit connection of these quantities with
physical parameters to model compliant contact with regularized friction. We
seek to model compliant contact as in Eq. (\ref{eq:compliant_model}),
parameterized by physical parameters: stiffness $k$ (in N/m) and
\textit{dissipation time scale} $\tau_d$ (in seconds). Therefore, users of this
model only need to provide these physical parameters and regularization is
computed from them. Notice this approach is different from the one in
\cite{bib:todorov2014}, where regularization is not used to model physical
compliance but rather to introduce a user tunable Baumgarte-style stabilization
to avoid constraint drift. 

% Dummy comment for Reviewable.

\subsection{Compliant Contact, Principle of Maximum Dissipation and Artifacts}
\label{sec:physical_intuition}

Dropping subscript $i$ for simplicity, we solve the projection problem in Eq.
(\ref{eq:y_projection}) analytically in Appendix
\ref{app:analytical_inverse_dynamics_derivations} for a regularization matrix of
the form $\vf{R} = \text{diag}([R_t, R_t, R_n])$
\begin{eqnarray}
	\bgamma &=& P_\mathcal{F}(\vf{y}) \label{eq:analytical_y_projection}\\
    &=&\begin{dcases}
	% Region I, stiction
	\vf{y} 
	% When we  have:
	& \text{stiction, } y_r \le \mu y_n\\
	%
	%
	% Region II, sliding.
	\begin{bmatrix}
		\mu\gamma_n\hat{\vf{t}}\\
		\frac{1}{1+\tilde\mu^2}\left(y_n + \hat\mu y_r\right)
	\end{bmatrix}
	% When we  have:
	& \text{sliding, } -\hat\mu y_r < y_n \leq \frac{y_r}{\mu}\\
	%
	%
	% Region III, no contact.
    \vf{0} & \text{no contact, } y_n < -\hat\mu y_r \end{dcases}\nonumber	
\end{eqnarray}
where $\vf{y}_t$ and $y_n$ are the tangential and normal components of $\vf{y}$,
$y_r=\Vert\vf{y}_t\Vert$ is the radial component, and
$\hat{\vf{t}}=\vf{y}_t/y_r$ is the unit tangent vector. We also define the
coefficients $\tilde\mu=\mu\,(R_t/R_n)^{1/2}$ and $\hat\mu=\mu\,R_t/R_n$ that
result from the \textit{warping} introduced by the metric $\vf{R}$.

Our compliant model of contact is defined by
\begin{eqnarray*}
	\hat{\vf{v}}_c &=&
	\begin{bmatrix}
		0\\
		0\\
		\hat{v}_n \end{bmatrix}\nonumber,\\
	\hat{v}_n &=& -\frac{\phi_0}{\delta t+\tau_d},
\end{eqnarray*}
where $\phi_0$ is the previous step signed distance reported by the geometry engine. The normal direction
regularization parameters is taken as $R_n^{-1} = \delta t k(\delta t+\tau_d)$.
To gain physical insight into our model, we substitute
$\vf{y}=-\vf{R}^{-1}(\vf{v}_c - \hat{\vf{v}}_c)$ into Eq.
(\ref{eq:analytical_y_projection}) to obtain 
\begin{align*}
	&\bgamma(\vf{v}_c) = P_\mathcal{F}(\vf{y}(\vf{v}_c))\\
&=\begin{dcases}
	% Region I, stiction
	\begin{bmatrix}
		-\vf{v}_t/R_t\\
		-\delta t\,k\,(\phi + \tau_d\,v_n)
	\end{bmatrix}
	% When we  have: y_r < \mu y_n
	& \text{stiction, } \\
	%
	%
	% Region II, sliding.
	\begin{bmatrix}
		\mu\gamma_n\hat{\vf{t}}\\
		-\frac{\delta t}{1+\tilde\mu^2}k\left(\phi-(\delta
		t+\tau_d)\mu\Vert\vf{v}_t\Vert + \tau_d\,v_n \right)
	\end{bmatrix}
	% When we  have:  -\mu \frac{R_t}{R_n} y_r < y_n \leq \frac{y_r}{\mu}
	& \text{sliding, }\\
	%
	%
	% Region III, no contact.  y_n \leq -\mu \frac{R_t}{R_n} y_r
    \vf{0} & \text{no contact, } \end{dcases}\nonumber	
\end{align*}
where $\phi= \phi_0 + \delta t\,v_n$ approximates the signed distance function
at the next time step.

Let us now analyze the resulting impulses from this
model.

\textbf{Friction Impulses}. We see that friction impulses behave exactly as a model
of regularized friction
\begin{equation}
	\bgamma_t = \min\left(\frac{\Vert\vf{v}_t\Vert}{R_t}, \mu\gamma_n\right)\hat{\vf{t}},
	\label{eq:regularized_friction}
\end{equation}
with $\bgamma_t$ linear with the (very small) slip velocity during stiction and
with the maximum value given by $\mu\gamma_n$, effectively modeling Coulomb's
friction. Notice that to better model stiction, we are interested in small
values of $R_t$. We discuss our parameterization of $R_t$ in Section
\ref{sec:conditioning}. Moreover, since $\hat{\vf{t}} =
\vf{y}_t/\Vert\vf{y}_t\Vert = -\vf{v}_t/\Vert\vf{v}_t\Vert$, friction impulses
oppose sliding and therefore satisfy the principle of maximum dissipation.

\textbf{Normal impulses}. We observe that in stiction, we recover the compliant
model given by Eq. (\ref{eq:compliant_model}), as desired. In the sliding
region, however, we see that the convex approximation introduces unphysical
artifacts. 

Firstly, the factor $1+\tilde{\mu}^2$ models an effective stiffness
$k_\text{eff}=k/(1+\tilde{\mu}^2)$ different from the physical value. Therefore
to accurately model compliance during sliding we must satisfy the condition
$\tilde\mu=\mu\,(R_t/R_n)^{1/2} \approx 0$ or, equivalently, $R_t \ll R_n$.
Section \ref{sec:conditioning} introduces a parameterization of $R_t$ that
satisfies this condition.

Secondly, we see that the slip velocity $\vf{v}_t$ unphysically couples into the
normal impulses as $\gamma_n=-\delta t k (\phi_\text{eff} + \tau_d\,v_n)$ with
an \textit{effective} signed distance $\phi_\text{eff} = \phi-(\delta
t+\tau_d)\mu\Vert\vf{v}_t\Vert$. That is, we recover the dynamics of compliant
contact but with a spurious drift of magnitude $(\delta
t+\tau_d)\mu\Vert\vf{v}_t\Vert$. This is consistent with the formulation in
\cite{bib:anitescu2010} for rigid contact when $k\rightarrow \infty$ and
$\tau_d=0$, leading to an unphysical \textit{gliding effect} at a positive
distance $\phi=\delta t\mu\Vert\vf{v}_t\Vert$. Notice that the \textit{gliding}
goes away as $\delta t\rightarrow 0$ since the formulation converges to the
original contact problem \cite{bib:anitescu2006}. The effect of compliance is to
\textit{soften} this gliding effect. With finite stiffness, the normal impulse
when sliding goes to $-k(\phi-\tau_d\mu\Vert\vf{v}_t\Vert)-d\,v_n$ in the limit
to $\delta t\rightarrow 0$. This tells us that, unlike the rigid case, the
\textit{gliding} effect unfortunately does not go away as $\delta t\rightarrow
0$. It persists with a finite value that now depends on the dissipation rate,
$\phi\approx\tau_d\mu\Vert\vf{v}_t\Vert$.

We close this discussion by making the following remarks relevant to robotics
applications:
\begin{enumerate}
	\item We are mostly interested in the stiction regime, typically for
	grasping, locomotion, or rolling contact for mobile bases with wheels. This
	regime is precisely where the convex approximation does not introduce
	artifacts.
	\item Sliding usually happens with low velocities and therefore the term
	$\delta t\mu\Vert\vf{v}_t\Vert$ is negligible.
	\item For robotics applications, we are mostly interested in inelastic
	contact. We will see that this can be effectively modeled with
	$\tau_d\approx\delta t$ in Section \ref{sec:conditioning}.
	Therefore, in this regime, the term $\tau_d\mu\Vert\vf{v}_t\Vert$ also goes
	to zero as $\delta t\rightarrow 0$.
	\item We are definitely interested in the onset of sliding. This is captured
	by the approximation which properly models the Colulomb friction law.
\end{enumerate}

% Dummy comment for Reviewable.

\subsection{Conditioning of the Problem}
\label{sec:conditioning}

Regularization parameters not only determine the physical model, but also affect
the robustness and performance of the SAP solver. Modeling near-rigid objects
and avoiding viscous drift during stiction require very small values of $R_t$
and $R_n$ that can lead to badly ill-conditioned problems. Under these
conditions, the Hessian of the system exhibits a large condition number, and
round-off errors can render the search direction of Newton iterations useless.
We show in this section how a judicious choice of the regularization parameters
leads to much better conditioned system of equations, without sacrificing
accuracy. This is demonstrated in Section \ref{sec:test_cases} with a variety of
tests cases.

\textbf{Near-Rigid Contact}. In our formulation rigid objects must be modeled as
\emph{near-rigid} using large stiffnesses. However, as mentioned above,
blindly choosing large values of stiffness can lead to ill-conditioned systems
of equations. Here, we propose a principled way to choose the stiffness
parameter when modeling near-rigid contact.

Consider the dynamics of a mass particle $m$ laying on the ground, with contact
stiffness $k$ and dissipation time scale $\tau_d$. When in contact, the dynamics
of this particle is described by the equations of a harmonic oscillator with
natural frequency $\omega_n^2 = k/m$, or period $T_n = 2\pi/\omega_n$, and
damping ratio $\zeta=\tau_d\omega_n/2$. We say the contact is \emph{near-rigid}
when $T_n \lesssim \delta t$ and the time step $\delta t$ cannot temporally
resolve the contact dynamics. In this \emph{near-rigid} regime, we use
compliance as a means to add a Baumgarte-like \emph{stabilization} to avoid
constraint drift, as similarly done in \cite{bib:todorov2011}. Choosing the time
scale of the contact to be $T_n = \beta \delta t$ with $\beta \le 1$, we model
inelastic contact with a dissipation that leads to a critically damped
oscillator, or $\zeta=1$. This dissipation is $\tau_d=2\zeta/\omega_n$, or in
terms of the time step,
\begin{equation*}
    \tau_d=\frac{\beta}{\pi}\delta t.
\end{equation*}

Using the harmonic oscillator equations, we can estimate the value of stiffness
from the frequency $\omega_n$ as $k=4\pi^2 m/(\beta^2 \delta t^2)$. Since
$\tau_d\approx\delta t$, $R_n^{-1} = \delta t k(\delta t+\tau_d) \approx \delta
t^2k$, and we estimate the regularization parameter as
\begin{equation*}
	R_n = \frac{\beta^2}{4\pi^2}\text{w},
\end{equation*}
where we define $\text{w}=1/m$.

It is useful to estimate the amount of penetration for a point mass resting on
the ground. In this case we have
\begin{align*}
	\phi = \frac{mg}{k} = \frac{\beta^2}{4\pi^2}m\text{w}g\delta t^2 = \frac{\beta^2}{4\pi^2}g\delta t^2,
\end{align*}
independent of mass. Taking $\beta=1.0$ and Earth's gravitational constant, a
typical simulation time step of $\delta t=10^{-3}~\text{s}$ leads to
$\phi\approx 2.5\times 10^{-7}~\text{m}$, and a large simulation time step of
$\delta t=10^{-2}~\text{s}$ leads to $\phi\approx 2.5\times 10^{-5}~\text{m}$,
well within acceptable bounds to consider a body rigid for typical robotics
applications.

For a general multibody system, we define the per-contact effective mass as
$\text{w}_i=\Vert\mathbf{W}_{ii}\Vert_\text{rms}=\Vert\mathbf{W}_{ii}\Vert/3$
where $\mathbf{W}_{ii}$ is the $3\times 3$ diagonal block of the Delassus
operator $\mathbf{W}=\mf{J}\mf{M}^{-1}\mf{J}^T$ for the $i$-th contact.
Explicitly forming the Delassus operator is an expensive operation. Instead we
use an $\mathcal{O}(n)$ approximation. Given contact $i$ involving trees $t_1$
and $t_2$, we form the approximation $\mathbf{W}_{ii}\approx\mf{J}_{i
t_1}\mf{M}_{t_1}^{-1}\mf{J}_{i t_1}^T+\mf{J}_{i t_2}\mf{M}_{t_2}^{-1}\mf{J}_{i
t_2}^T$. Finally, we compute the
regularization parameter in the normal direction as
\begin{eqnarray}
    R_n = \max\left(\frac{\beta^2}{4\pi^2}\Vert\mathbf{W}_{ii}\Vert_\text{rms},
    \frac{1}{\delta t k(\delta t+\tau_d)}\right)
    \label{eq:normal_regularization}.
\end{eqnarray}

With this strategy, our model automatically switches between modeling compliant
contact with stiffness $k$ when the time step $\delta t$ can resolve the
temporal dynamics of the contact, and using stabilization to model near-rigid
contact with the amount of stabilization controlled by parameter $\beta$. In all
of our simulations, we use $\beta=1.0$.

\textbf{Stiction}. Given that our model regularizes friction, we are interested
in estimating a bound on the slip velocity at stiction. We propose the following
regularization for friction
\begin{equation}
    R_t = \sigma \text{w},
    \label{eq:tangential_regularization}
\end{equation}
where $\sigma$ is a dimensionless parameter.

To understand the effect of $\sigma$ in the approximation of stiction, we
consider once again a point of mass $m$ in contact with the ground under
gravity, for which $\text{w}\approx 1/m$. We push the particle with a horizontal
force of magnitude $F=\mu\gamma_n$ so that friction is right at the boundary of
the friction cone and the slip velocity due to regularization, $v_s$, is
maximized. Then in stiction, we have
\begin{equation*}
    \|\bgamma_t\| = \frac{v_s}{R_t} = \mu m g \delta t.
\end{equation*}
Using our proposed regularization in Eq. (\ref{eq:tangential_regularization}),
we find the maximum slip velocity
\begin{equation}
    v_s \approx \mu\sigma g \delta t,
    \label{eq:slip_estimation}
\end{equation}
independent of the mass and linear with the time step size. Even though the
friction coefficient $\mu$ can take any non-negative value, most often in
practical applications $\mu < 1$. Values on the order of 1 are in fact
considered as large friction values. Therefore, for this analysis we consider
$\mu\approx 1$. In all of our simulations, we use $\sigma=10^{-3}$. With Earth's
gravitational constant, a typical simulation with time step of $\delta
t=10^{-3}~\text{s}$ leads to a stiction velocity of $v_s\approx
10^{-5}\text{m}/\text{s}$, and with a large step of $\delta t=10^{-2}~\text{s}$,
$v_s\approx 10^{-4}\text{ m}/\text{s}$. Smaller friction coefficients lead to
even tighter bounds. These values are well within acceptable bounds even for
simulation of grasping tasks, which is significantly more demanding than
simulation for other robotic applications, see Section \ref{sec:test_cases}.

\textbf{Sliding Soft Contact}. As we discussed in Section
\ref{sec:physical_intuition}, we require $R_t/R_n\ll 1$ so that we model
compliance accurately during sliding. Now, in the \emph{near-rigid} contact
regime, the condition $R_t/R_n\ll 1$ is no longer required since in this regime
regularization is used for stabilization. Therefore, we only need to verify this condition in the \emph{soft contact}
regime, when time step $\delta t$ can properly resolve the contact dynamics,
i.e. according to our criteria, when $\delta t < T_n$. In this regime,
$R_n^{-1}\approx \delta t^2k$, and using Eq.
(\ref{eq:tangential_regularization}) we have
\begin{equation*}
    \frac{R_t}{R_n}\approx \sigma \delta t^2 \omega_n^2=4\pi^2\sigma\left(\frac{\delta t}{T_n}\right)^2
    \lesssim 4\pi^2\sigma
\end{equation*}
where in the last inequality we used the assumption that we are in the soft
regime where $\delta t < T_n$. Since $\sigma \ll 1$ and in particular we use
$\sigma=10^{-3}$ in all of our simulations, we see that $R_t/R_n \ll 1$.
Moreover, $R_t/R_n$ goes to zero quadratically with $\delta t/T_n$ as the time
step is reduced and the dynamics of the compliance is better resolved in time.

Summarizing, we have shown that our choice of regularization parameters enjoys
the following properties
\begin{enumerate}
    \item Users only provide physical parameters; contact stiffness $k$,
    dissipation time scale $\tau_d$, and friction coefficient $\mu$. There is no
    need for users to tweak solver parameters.
    \item In the \emph{near-rigid} limit, our regularization in Eq.
    (\ref{eq:normal_regularization}) automatically switches the method to model
    rigid contact with constraint stabilization to avoid excessively large
    stiffness parameters and the consequent ill-conditioning of the system.
    \item Frictional regularization is parameterized by a single dimensionless
    parameter $\sigma$. We estimate a bound for the slip velocity during
    stiction to be $v_s \approx \mu \sigma \delta t g$. For $\sigma=10^{-3}$,
    the slip during stiction is well within acceptable bounds for robotics
    applications.
    \item We show that $R_t/R_n \ll 1$ when $\delta t$ can resolve the dynamics
    of the compliant contact, as required to accurately model compliance during
    sliding.
\end{enumerate}

\section{Test Cases}
\label{sec:test_cases}

We evaluate the robustness, accuracy, and performance of our method in a number
of simulation tests. All simulations are carried out in a system with 24 2.2 GHz
Intel Xeon cores (E5-2650 v4) and 128 GB of RAM, running Linux. However, all of
our tests are run in a single thread.

For all of our simulations, unless otherwise specified, our model uses
$\beta=1.0$ and $\sigma=10^{-3}$ for the regularization parameters in Eq.
(\ref{eq:normal_regularization}) and Eq. (\ref{eq:tangential_regularization}),
respectively.

\subsection{Performance Comparisons Against Other Solvers}
\label{sec:about_solvers}

We evaluate commercial software Gurobi, considered an industry standard, to
solve our primal formulation \eqref{eq:primal_regularized}. As an open source
option, we evaluate the Geodesic interior-point method (IPM) from
\cite{bib:permenter2020}. Geodesic IPMs, in contrast with primal-dual IPMs, do
not apply Newton's method to the central-path conditions directly. Instead, they
use geodesic curves that satisfy the complementarity slackness condition. Since
the Geodesic IPM and SAP use the same supernodal linear algebra code described
in Section \ref{sec:problem_sparsity}, it is natural to compare their
performance.

For performance comparisons, we use the steady clock from the STL
\verb+std::chrono+ library to measure wall-clock time for SAP and Geodesic IPM.
For Gurobi we access the \verb+Runtime+ property reported by Gurobi. Notice this
is somewhat unfair to SAP and Geodesic IPM since Gurobi's reported time does not
include the cost of the initial setup.

\subsection{Spring-Cylinder}
\label{sec:spring_cylinder}
\begin{figure}[!h]
	\centering
	\includegraphics[width=0.6\columnwidth]{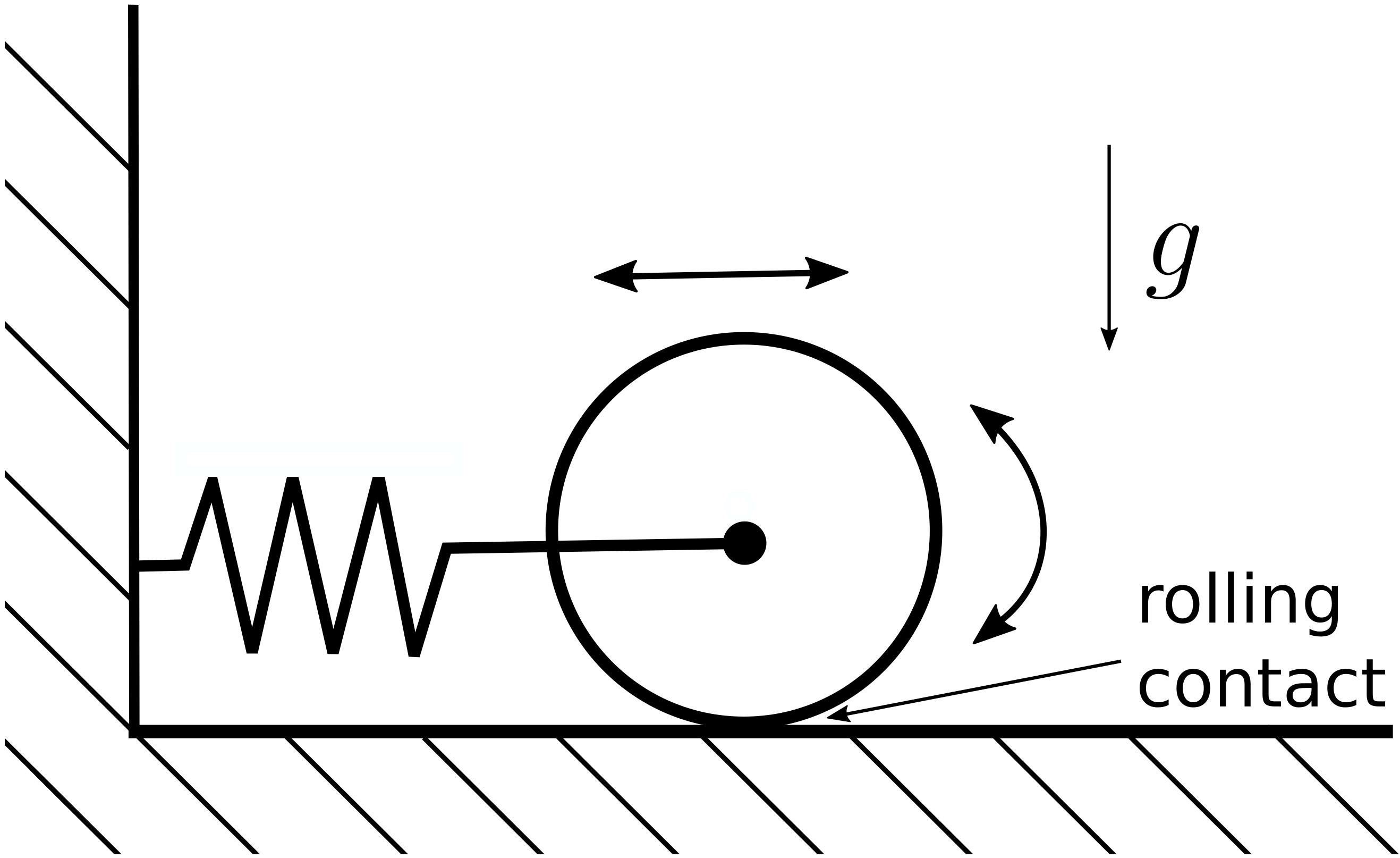}
	\caption{\label{fig:spring_cylinder} 
	Spring-Cylinder system. The cylinder can translate horizontally and rotate.
	Friction with the ground establishes a non-dissipative rolling contact.}
\end{figure}

We model the setup shown in Fig. \ref{fig:spring_cylinder}, consisting of a
cylinder of radius $R=0.05\text{ m}$ and mass $m=0.5\text{ kg}$ connected to a
wall to its left by a spring of stiffness $k_s=100\text{ N}/\text{m}$. While the
cylinder is free to rotate and translate in the plane, the ground constrains the
cylinder's motion in the vertical direction. The contact stiffness is
$k=10^{4}\text{ N}/\text{m}$ and the dissipation time scale is
$\tau_d=0.02\text{ s}$. The cylinder is initially placed with zero velocity at
$x_0=0.1\text{ m}$ to the right of the spring's resting position, and it is then
set free.

\begin{figure}[!h]
    \centering
    %trim={<left> <lower> <right> <upper>}
    \adjincludegraphics[width=0.49\columnwidth,trim={0 0 {0.05\width} 0},clip]{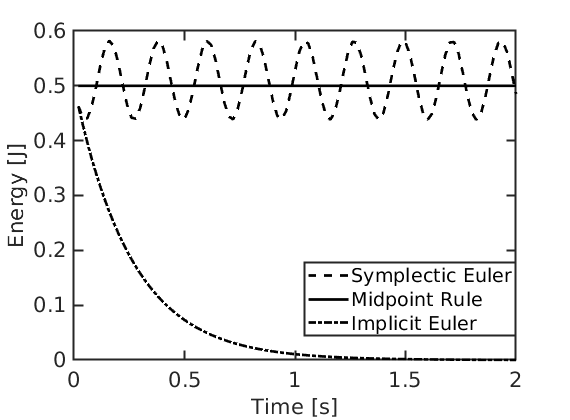}
    \adjincludegraphics[width=0.49\columnwidth,trim={0 0 {0.05\width} 0},clip]{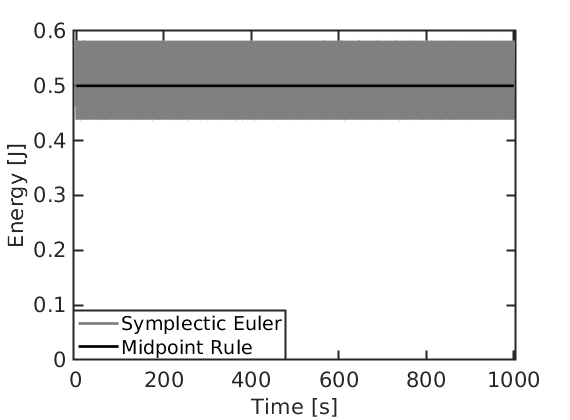}    
    \caption{\label{fig:frictionless_spring_cylinder_energy} 
    Total mechanical energy for the frictionless spring-cylinder system in the
    first few periods of oscillation (left) and long term (right).}
\end{figure}

For reference, we first simulate this setup with frictionless contact, i.e.
with $\mu=0$. Without friction, the cylinder does not rotate and we effectively have
a spring-mass system with natural frequency $\omega_n=\sqrt{k_s/m}$. We use a
rather coarse time step of $\delta t=0.02\text{ s}$, discretizing each period of
oscillation with about $22$ steps. Figure
\ref{fig:frictionless_spring_cylinder_energy} shows the total mechanical energy
as a function of time computed using three different schemes; symplectic Euler,
midpoint rule, and implicit Euler. The amount of numerical dissipation
introduced by the implicit Euler scheme dissipates the initial energy in just a
few periods of oscillation. For the symplectic Euler scheme, we observe in Fig.
\ref{fig:frictionless_spring_cylinder_energy} that, while the energy is not
conserved, it stays bounded, within a band 28\% peak-to-peak wide. The figure
also confirms that the second order midpoint scheme conserves energy exactly.
These are well known theoretical properties of these integration schemes when
applied to the spring-mass system.

We now focus our attention to a case with frictional contact using $\mu=1$. As
we release the cylinder from its initial position at $x_0=0.1\text{ m}$,
friction with the ground establishes a rolling contact, and the system sets into
periodic motion. Since now kinetic energy is split into translational and
rotational components, the rolling cylinder behaves as a spring-mass system with
an effective mass $m_\text{eff}=m+I_o/R^2$, with $I_o$ the rotational inertia of
the cylinder about its center. Therefore the frequency of oscillation is slower,
and the same time step, $\delta t=0.02\text{ s}$, now discretizes one period of
oscillation with about 27 steps.

\begin{figure}[!h]
    \centering
    %trim={<left> <lower> <right> <upper>}
    \adjincludegraphics[width=0.49\columnwidth,trim={0 0 {0.05\width} 0},clip]{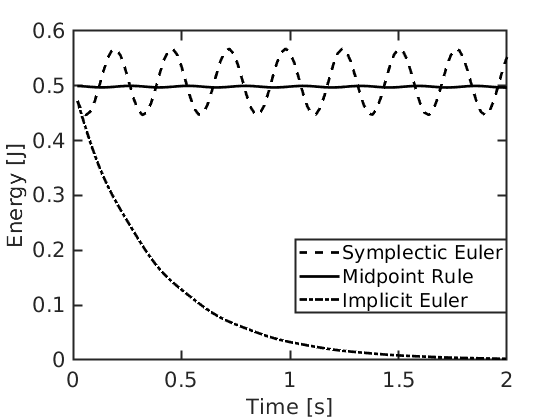}
    \adjincludegraphics[width=0.49\columnwidth,trim={0 0 {0.05\width} 0},clip]{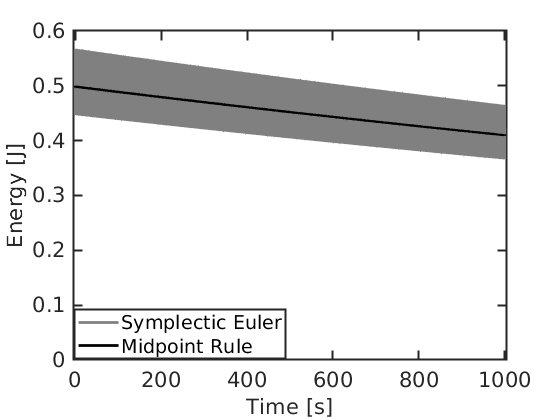}    
    \caption{\label{fig:spring_cylinder_energy} 
    Total mechanical energy for the spring-cylinder system with friction $\mu=1$
    in the first few periods of oscillation (left) and long term (right).}
\end{figure}

Total energy is shown in Fig. \ref{fig:spring_cylinder_energy}. Solutions
computed with the implicit Euler and the symplectic Euler scheme show similar
trends to those in the frictionless case. The midpoint rule does not conserve
energy exactly but it does significantly better, with a peak-to-peak variation
of only 0.16\%. While the ideal rolling contact does not dissipate energy, the
regularized model of friction does dissipate energy given the slip velocity is
never exactly zero, though small in the order of $\sim\sigma\mu\delta t g$ as
shown in Section \ref{sec:physical_intuition}. The symplectic Euler scheme and
the midpoint rule take $10$ minutes of simulated time and about $1000$
oscillations to dissipate 10\% of the total energy (Fig.
\ref{fig:spring_cylinder_energy}, right). This level of numerical dissipation is
remarkably low, considering that real mechanical systems often introduce several
sources of dissipation.

\begin{figure}[!h]
	\centering
	\includegraphics[width=0.7\columnwidth]{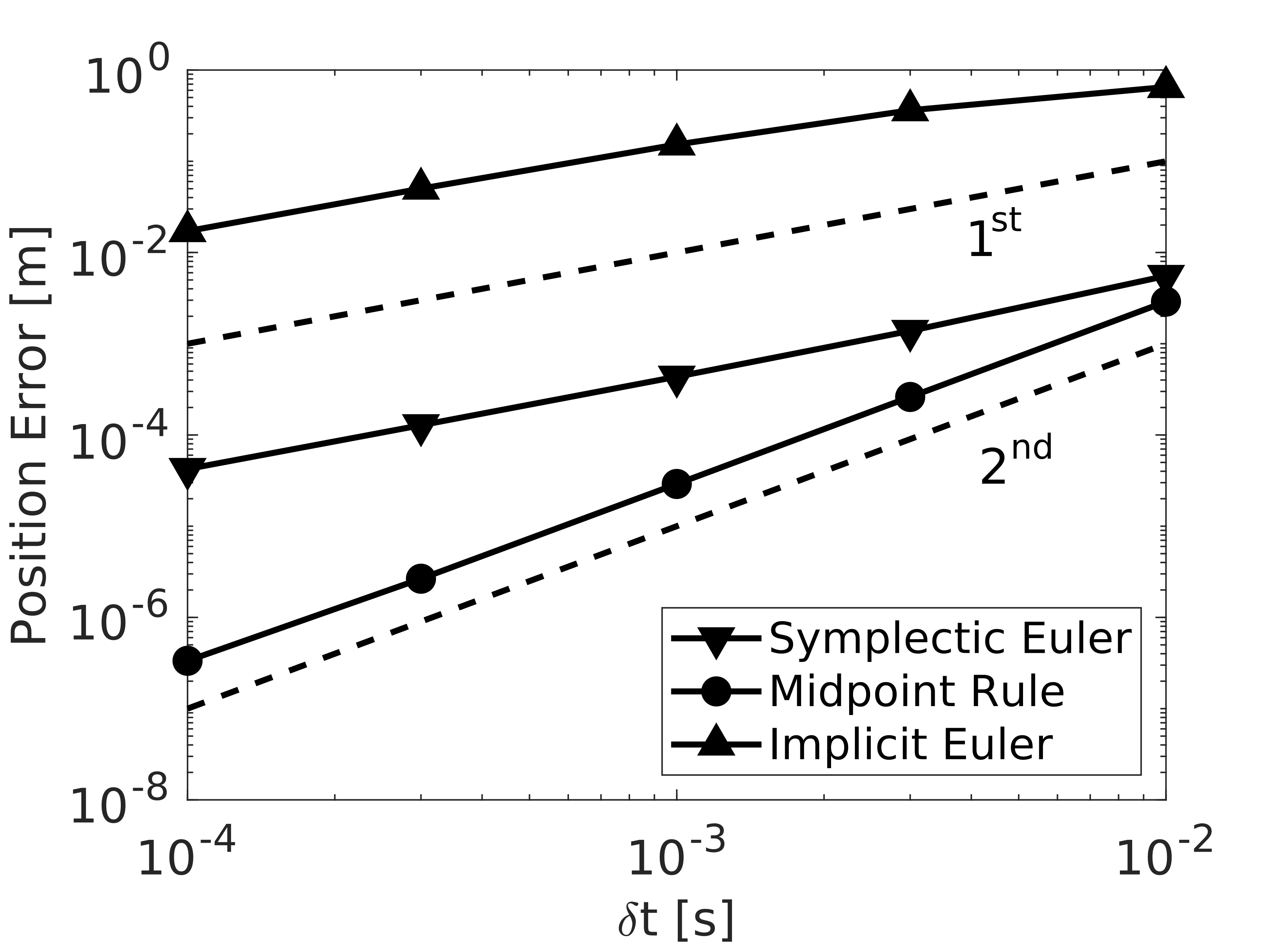}
	\caption{\label{fig:spring_cylinder_position_error} 
	Position error as a function of time step for the spring-cylinder system
	with friction. First and second order references are shown with dashed
	lines.}
\end{figure}

To study the order of accuracy of our approach, we define the $L^2$-norm
position error as
\begin{equation*}
    e_q = \left(\frac{1}{T}\int_0^T dt(x(t)-x_e(t))^2\right)^{1/2}
\end{equation*}
where $x_e(t)$ is the known exact solution. We simulate for $T=5\text{s}$, about
10 periods of oscillation. Figure \ref{fig:spring_cylinder_position_error} shows
the position error as a function of the time step. We see that even with
frictional contact, the two-stage approach with the midpoint rule achieves
second order accuracy. Both the implicit Euler and the symplectic Euler
scheme are first order, though the error is significantly smaller when using
the symplectic Euler scheme.

\subsection{Clutter}
\label{sec:clutter}
\begin{figure}[t]
    \centering
    %trim={<left> <lower> <right> <upper>}
    \adjincludegraphics[width=0.45\columnwidth,trim={0 {0.05\height} 0 0},clip]{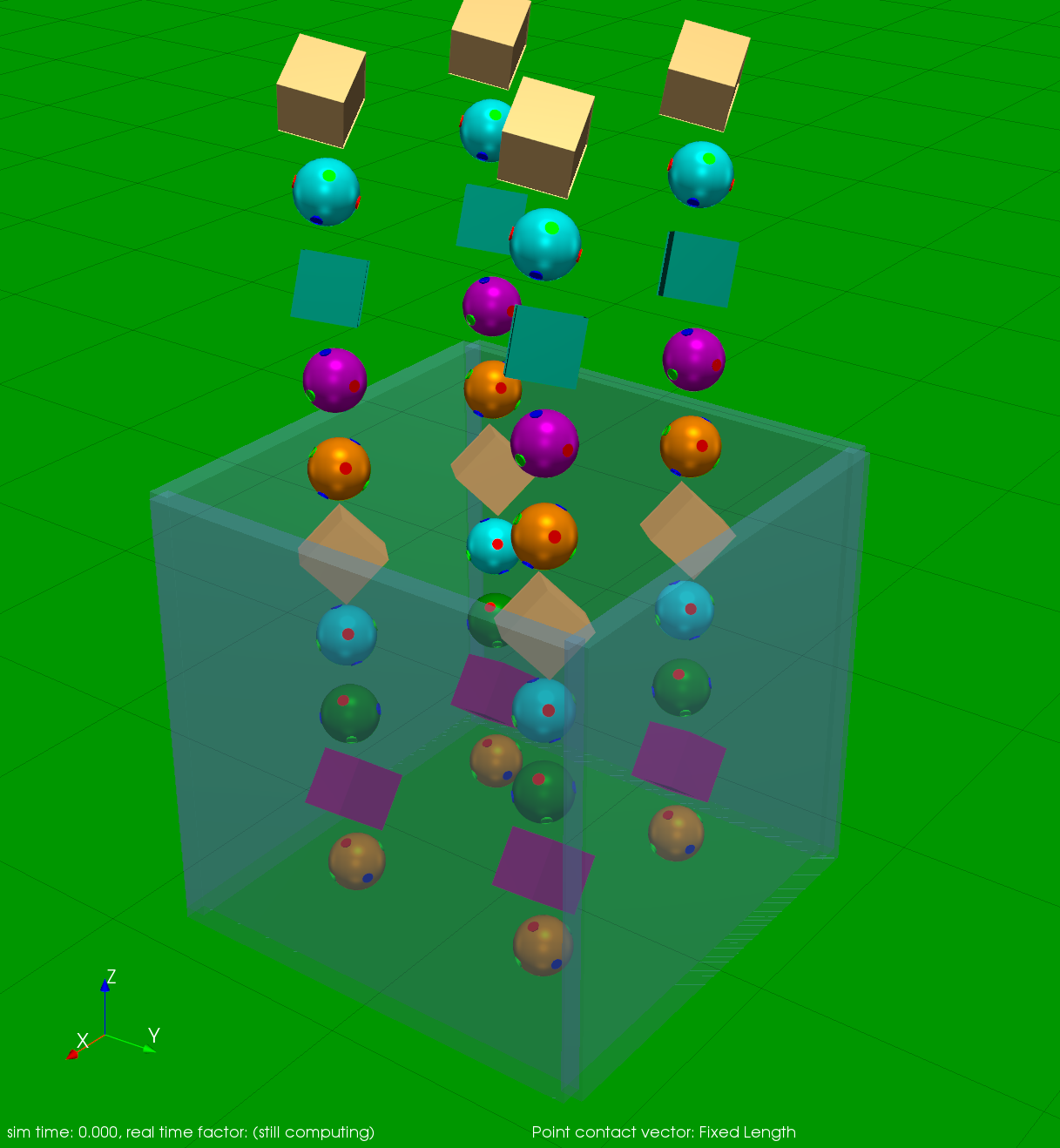}
    \adjincludegraphics[width=0.45\columnwidth,trim={0 {0.05\height} 0
    0},clip]{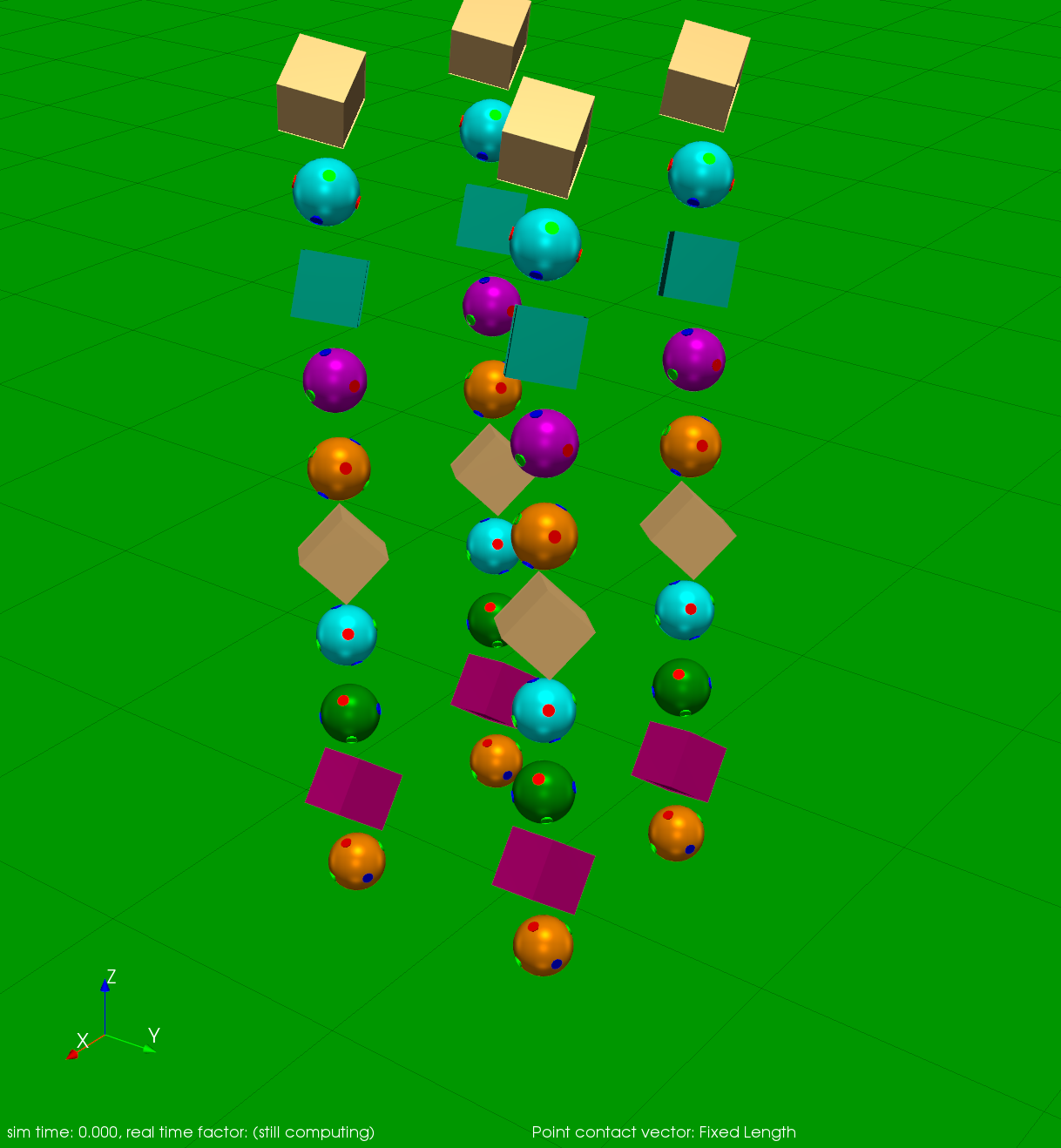}\\
    \vspace{0.1cm}
    \adjincludegraphics[width=0.45\columnwidth,trim={0 {0.05\height} 0 0},clip]{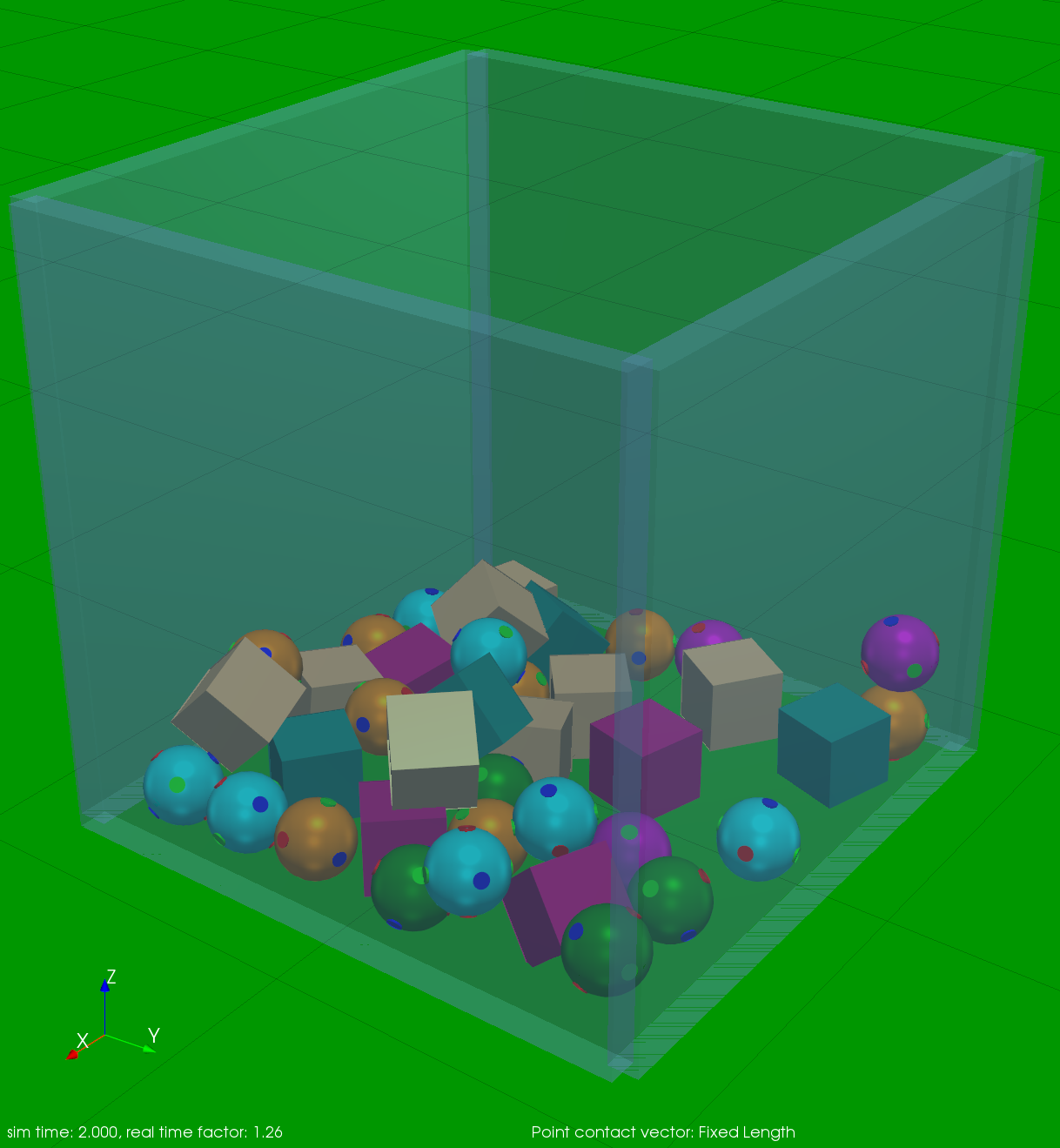}
    \adjincludegraphics[width=0.45\columnwidth,trim={0 {0.05\height} 0 0},clip]{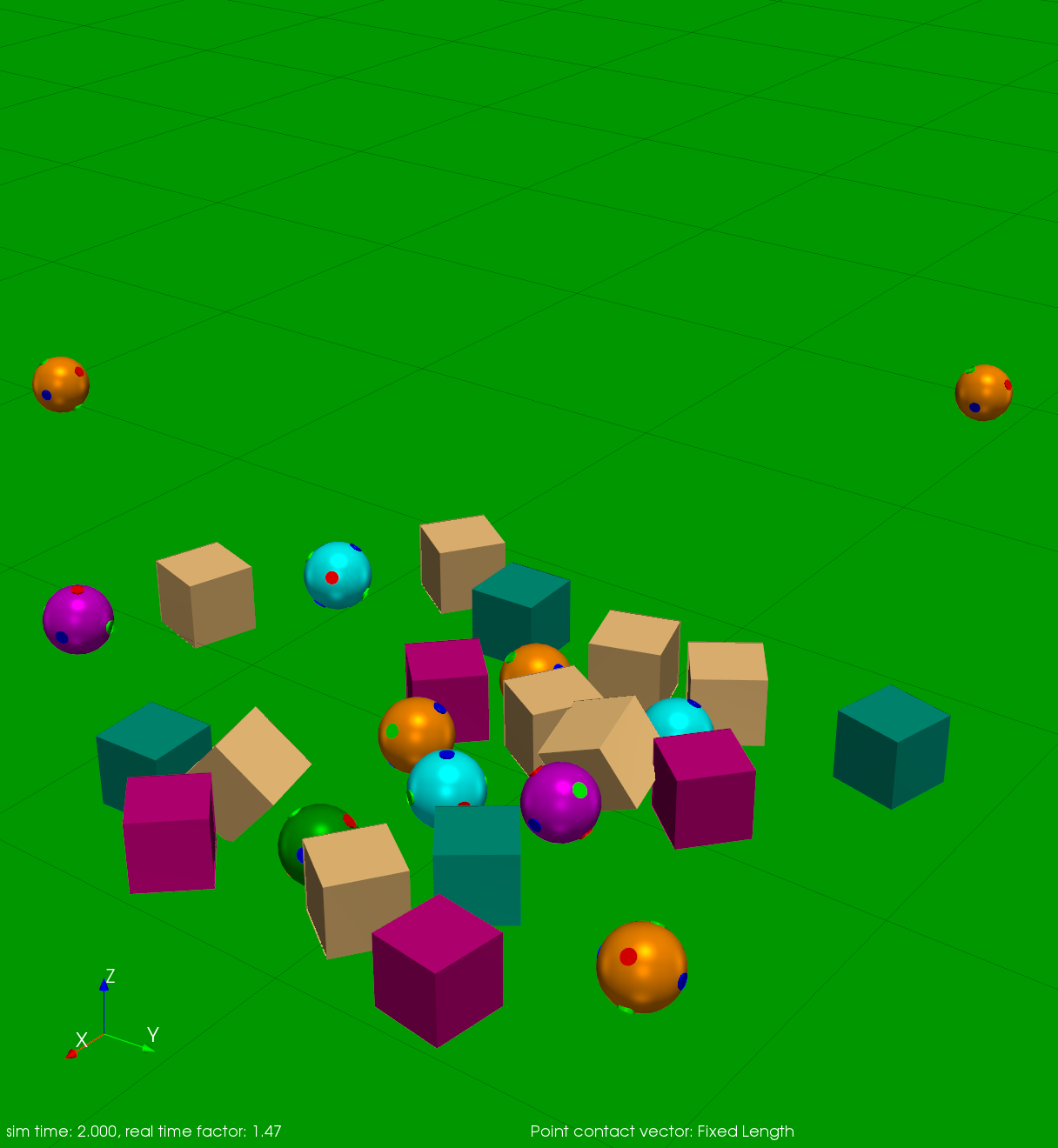}
    \caption{Initial conditions (top) and an intermediate configuration after
    $2$ seconds of simulated time (bottom) for the clutter setup with (left) and
    without (right) walls. Many of the spheres in the configuration with no
    walls roll outside the frame in the intermediate configuration.}
    \label{fig:clutter_snapshots}
\end{figure}

Objects are dropped into an $80~\text{cm}\times80~\text{cm}\times80~\text{cm}$
container in four different columns with the same number of objects in each (see
Fig. \ref{fig:clutter_snapshots}). Each column consists of an arbitrary
assortment of spheres of radius $5~\text{cm}$ and boxes with sides of
$10~\text{cm}$ in length. With a density of $1000\text{kg}/\text{m}^3$, spheres
have a mass of $0.524\text{ kg}$ and boxes have a mass of $1.0\text{ kg}$. We
set a very high contact stiffness of $k=10^{12}\text{ N}/\text{m}$ so that the
model is in the \emph{near-rigid} regime. The dissipation time scale is set to
equal the time step and the friction coefficient of all surfaces is $\mu=1.0$.

\begin{figure}[!h]
	\centering
    %trim={<left> <lower> <right> <upper>}
    \adjincludegraphics[width=0.49\columnwidth,trim={0 0 0 0},clip]{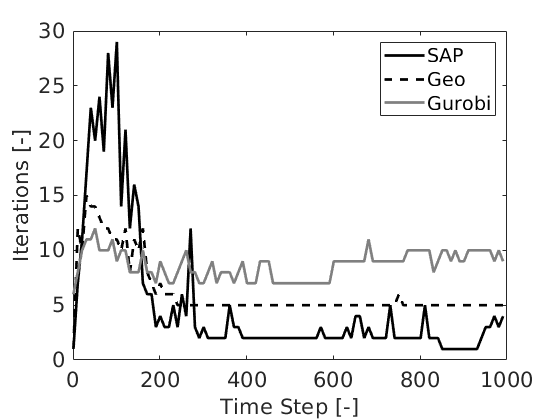}
    \adjincludegraphics[width=0.49\columnwidth,trim={0 0 0 0},clip]{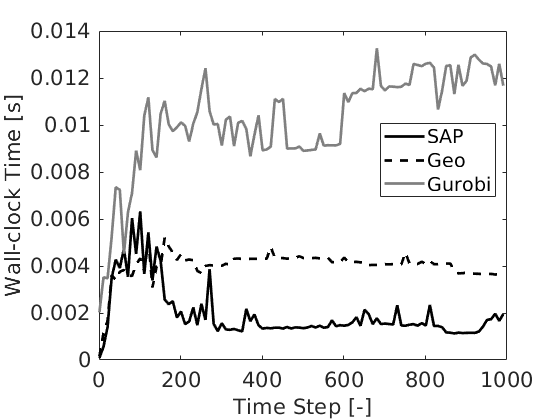}    
	\caption{\label{fig:clutter_w_walls_nb40} 
	Iterations and wall-clock time per time step for the clutter case with 40
	bodies and with walls.}
\end{figure}

We first run our simulations with 10 bodies per column for a total of 40 bodies.
We simulate 10 seconds using time steps of size $\delta t = 10\text{ ms}$.
Number of solver iterations and wall-clock time per time step are reported in
Fig. \ref{fig:clutter_w_walls_nb40}. We observe that SAP needs to perform a
larger number of iterations during the very energetic initial transient. As the
system reaches a steady state, however, SAP warm starts very effectively,
performing only about $3$ iterations per time step. Even though SAP necessities
a larger number of iterations to converge than Geodesic IPM during this initial
transient, the wall-clock time per time step is very similar. Unlike SAP and
Geodesic IPM that benefit from warm start, Gurobi performs about $9$
iterations per time step in both the initial transient and the steady state.

\begin{figure}[!h]
	\centering
    \includegraphics[height=0.34\columnwidth]{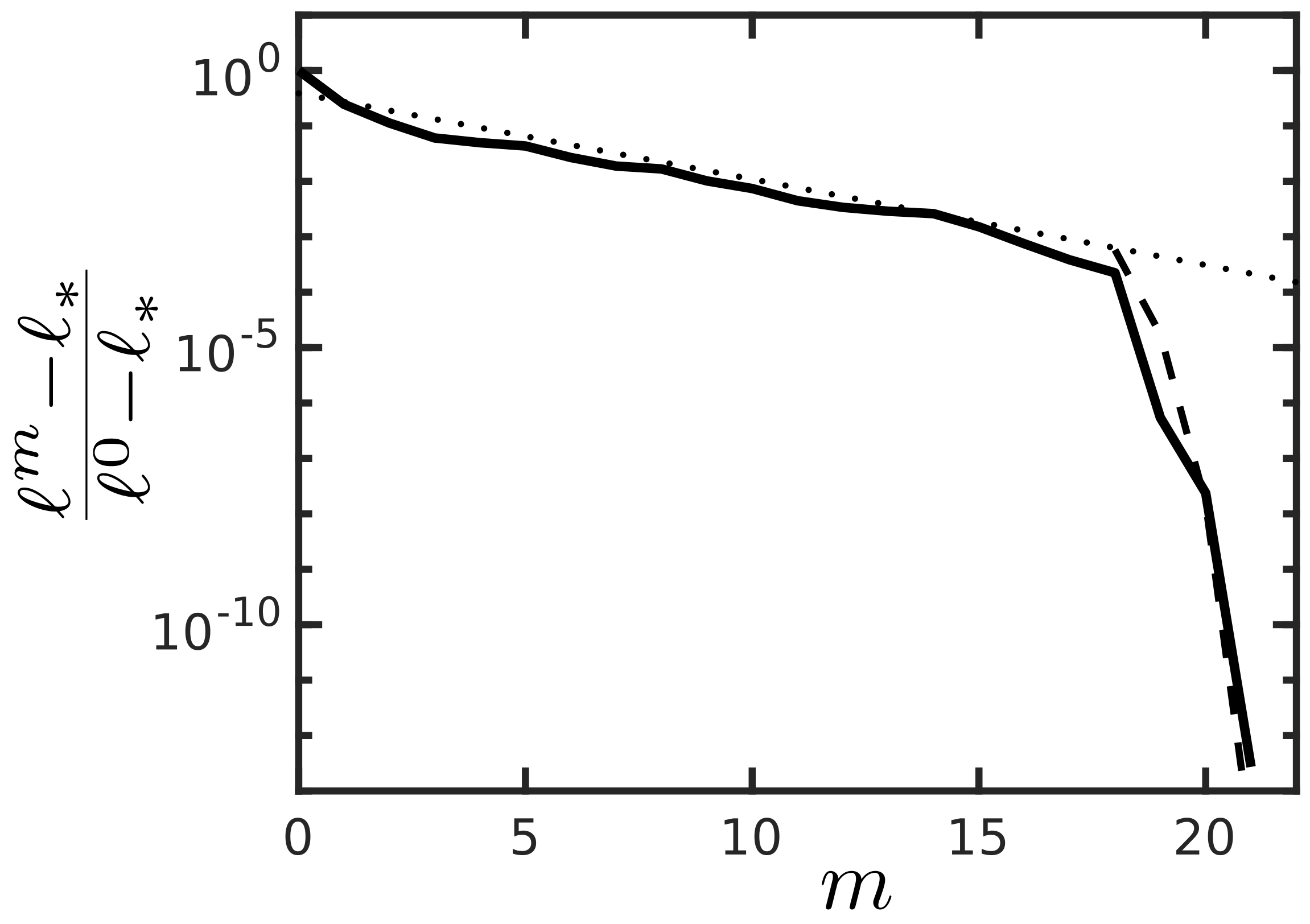}
	\includegraphics[height=0.34\columnwidth]{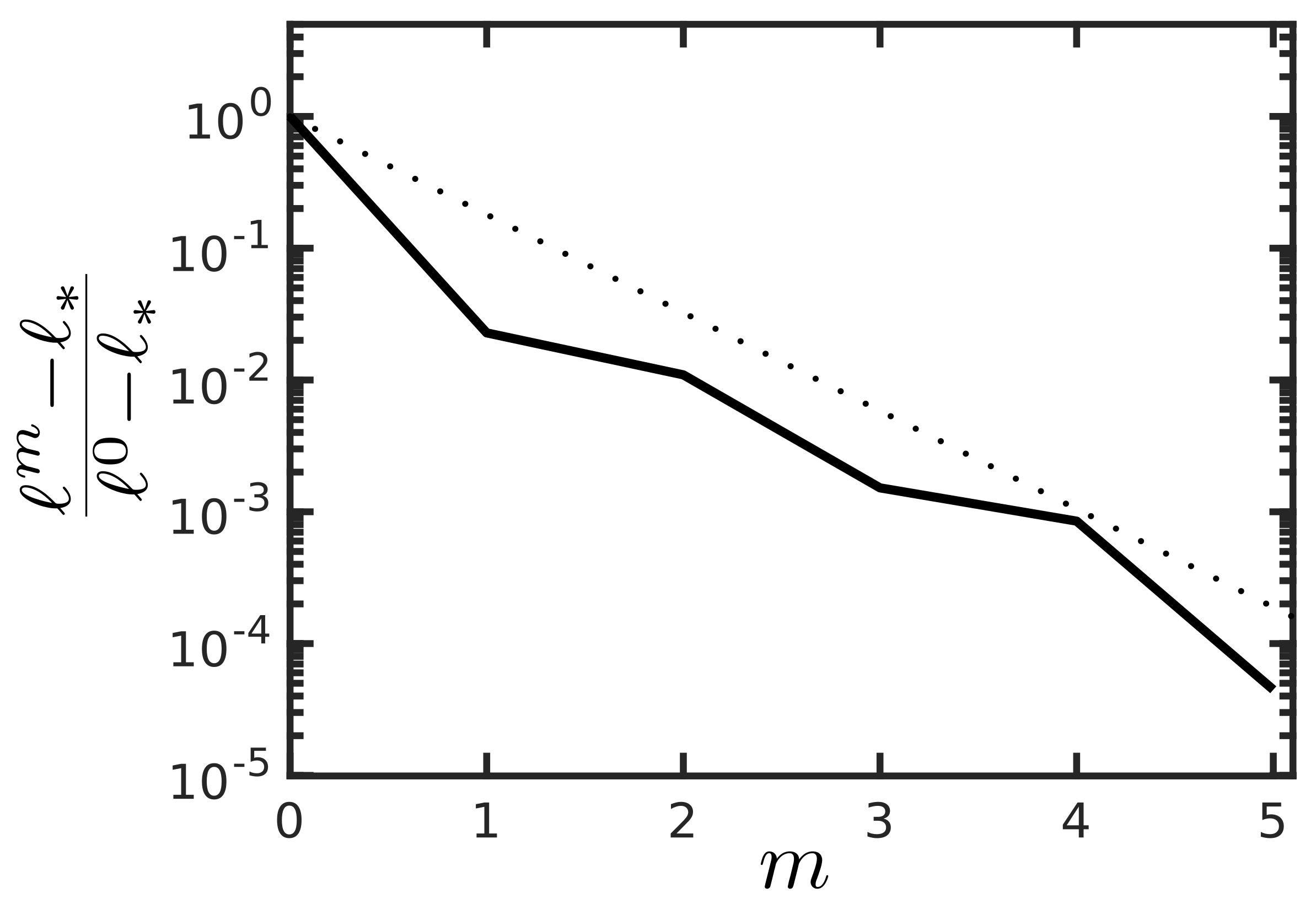}    
	\caption{\label{fig:clutter_line_search} 
	Cost as a function of Newton iterations for step 60 (left) and for step 520
	(right) using SAP. The cost decreases monotonically. Reference lines are
	shown for linear convergence (dotted) and quadratic convergence (dashed).}
\end{figure}

Figure \ref{fig:clutter_line_search} shows two examples of convergence history.
We denote with $\ell^0$ the cost evaluated at the initial guess, the previous
time step velocity. With $\ell_*$ we denote the optimal cost, which we
approximate with its value from the last iteration. At step 60 during the
initial transient for which SAP requires 21 iterations to converge, we observe
that the algorithm reaches quadratic convergence after an initial linear
convergence transient, matching theoretical predictions
(Appendix~\ref{app:sap_converge}). At step 520, past the initial energetic
transient, SAP exhibits linear convergence and satisfies the convergence
criteria within 5 iterations.

\subsubsection{Scalability}

\begin{figure}[!h]
	\centering
	\includegraphics[width=0.7\columnwidth]{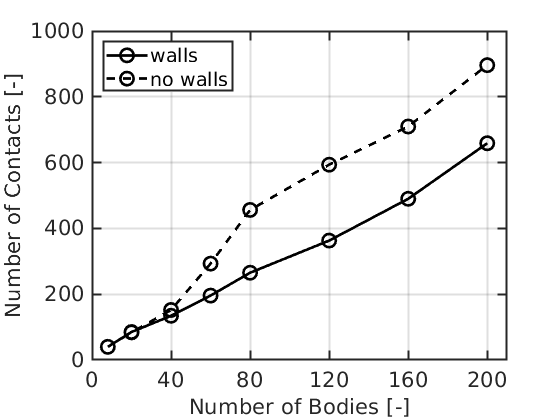}
	\caption{\label{fig:clutter_num_contats} 
	Total number of contacts with objects in steady state at the end of the
	simulation for setups with and without walls.}
\end{figure}

We evaluate the scalability of SAP by varying the number of objects in the
clutter. We study the case with and without walls (see Fig.
\ref{fig:clutter_snapshots}) as this variation leads to very different contact
configurations and sparsity patterns. The size of the problems can be
appreciated in Fig. \ref{fig:clutter_num_contats} showing the number of contact
constraints at the end of the simulation when objects are in steady state. We
observe a larger number of contacts for the configuration without walls since in
this configuration many of the boxes spread over the ground and lay flat on one
of their faces, leading to multi-contact configurations (see Fig.
\ref{fig:clutter_snapshots}).

\begin{figure}[!h]
	\centering
    %trim={<left> <lower> <right> <upper>}
    \adjincludegraphics[width=0.49\columnwidth,trim={{0.02\width} 0 {0.05\width} 0},clip]{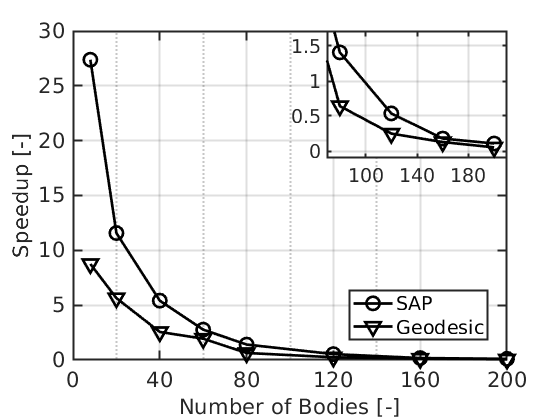}
    \adjincludegraphics[width=0.49\columnwidth,trim={{0.02\width} 0 {0.05\width} 0},clip]{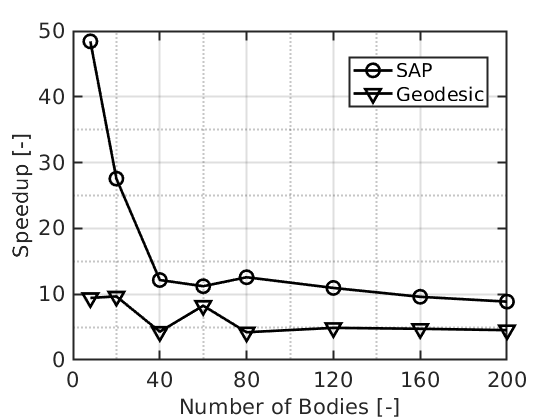}
	\caption{\label{fig:clutter_speedup} 
	Speedup against Gurobi for the configuration with walls (left) and without
	walls (right).}
\end{figure}

We define the \emph{speedup} against Gurobi as the ratio of the wall-clock time
spent by a solver to the wall-clock time reported by Gurobi. Figure
\ref{fig:clutter_speedup} shows the speedup for both SAP and Geodesic IPM in the
configuration with and without walls. The setup with walls is particularly
difficult given that objects are constrained to pile up, leading to a
configuration in which almost all objects are coupled with every other object by
frictional contact (see Fig. \ref{fig:clutter_snapshots}). For example, the motion
of an object at the bottom of the pile can lead to motion of another object far
on top of the pile. In contrast, the simulation with no walls leads to
\emph{islands} of objects that do not interact with each other.

In general, we observe two regimes. For problems with less than about 40 bodies,
SAP outperforms Gurobi significantly by up to a factor of 25 in the case with
walls and up to a factor of 50 with no walls. Beyond 80 bodies, Gurobi
outperforms both SAP and Geodesic IPM in the case with walls, but SAP is about
10 times faster for the case with no walls. Though SAP shows to be about twice
as fast as Geodesic IPM for most problem sizes, it can be five times faster for
small problems with 8 bodies or less.

It could be argued that these speedup results depend on the accuracy settings of
each solver. For a fair comparison, we define the dimensionless momentum error
as
\begin{equation}
	e_m = \frac{\Vert\tilde{\nabla}\ell_p\Vert}{\max(\Vert\tilde{\mf{p}}\Vert,\Vert\tilde{\mf{j}_c}\Vert)},
	\label{eq:momentum_error}
\end{equation}
using the scaled generalized momentum quantities in Eq.
(\ref{eq:scaled_momentum_quantities}). We also define the dimensionless
complementarity slackness error as
\begin{equation}
	e_\mu = \frac{1/n_c\sum_i|\bm{g}_i\cdot\bgamma_i|}{\ell_p}.
	\label{eq:slackness_error}
\end{equation}

\begin{figure}[!h]
	\centering
    %trim={<left> <lower> <right> <upper>}
    \adjincludegraphics[height=0.39\columnwidth,trim={0 0 {0.05\width} 0},clip]{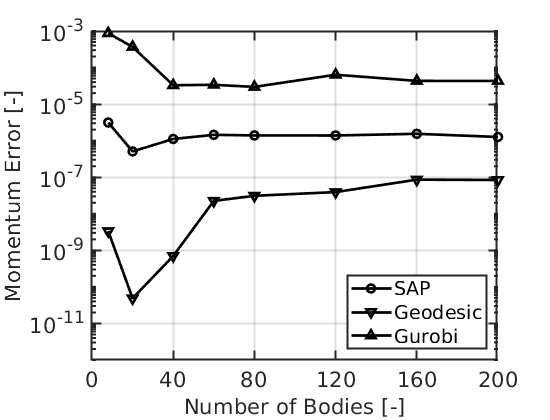}
	\adjincludegraphics[height=0.39\columnwidth,trim={{0.05\width} 0
	{0.05\width} 0},clip]{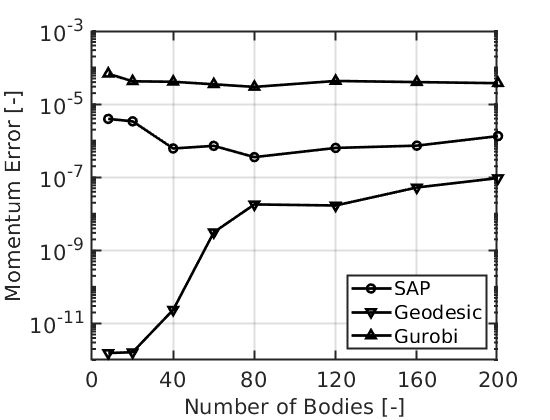}\\
    \adjincludegraphics[height=0.39\columnwidth,trim={0 0 {0.05\width} 0},clip]{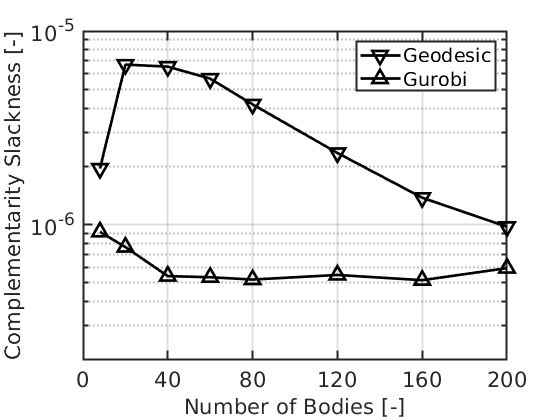}
    \adjincludegraphics[height=0.39\columnwidth,trim={{0.05\width} 0 {0.05\width} 0},clip]{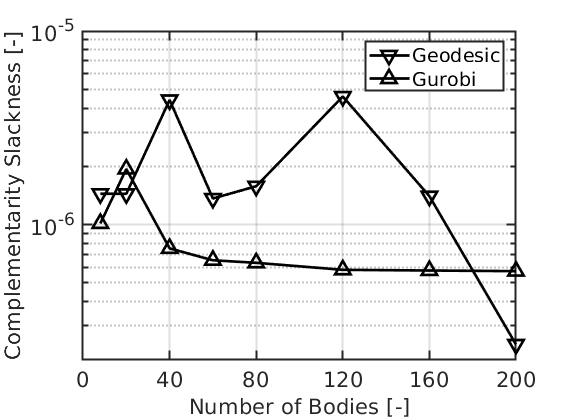}
	\caption{\label{fig:clutter_errors_w_wall} 
	Momentum balance error $e_m$ (top) and complementarity condition error
	$e_\mu$ (bottom) for the clutter case with walls (left) and without walls
	(right).}
\end{figure}

Figure \ref{fig:clutter_errors_w_wall} shows average values of $e_m$ and $e_\mu$
over all time steps. Since SAP satisfies the complementarity slackness exactly,
$e_\mu$ is not shown. We have verified this to be true within machine precision
for all simulated cases.

SAP's momentum error is below $10^{-5}$ as expected since this is the value used
for the termination condition. Similarly, the complementarity slackness is below
$10^{-5}$ for Geodesic IPM, since this is the value used for its own termination
condition. Gurobi does a good job at satisfying the complementarity slackness.
However, it is the solver with the largest error in the momentum equations, even
though both SAP and Geodesic IPM outperform Gurobi in most of the test cases.
These metrics demonstrate that when SAP and Geodesic IPM outperform Gurobi, it
is not at the expense of accuracy.

\subsubsection{Slip Parameter}
\begin{figure}[!h]
	\centering
	\includegraphics[width=0.7\columnwidth]{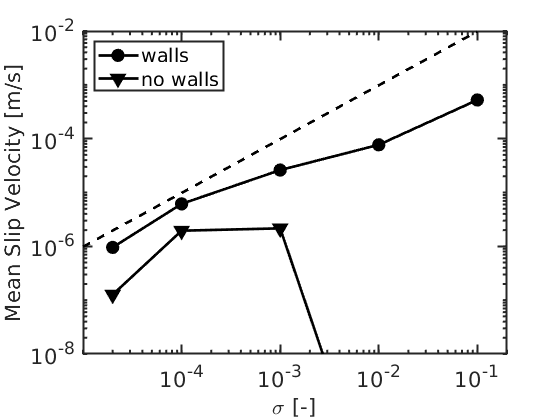}
	\caption{\label{fig:clutter_sigma_vt} 
	Mean slip velocity at the end of the simulation with objects at rest as a
	function of the slip parameter. The estimated bound $v_s = \sigma\mu\delta t
	g$ is shown in dashed lines.}
\end{figure}

We study the effect of the slip parameter $\sigma$ in Eq.
(\ref{eq:tangential_regularization}). We use $\delta t = 10\text{ ms}$ and
simulate with SAP 40 objects for 10 seconds to a steady state configuration. At
this steady state, we compute the mean slip velocity among all contacts, shown
in Fig. \ref{fig:clutter_sigma_vt} along with the estimated slip in Eq.
(\ref{eq:slip_estimation}), $v_s \approx\sigma\mu\delta t g$. We see that the
mean slip velocity remains below the estimated slip as expected in a static
configuration with objects in stiction. In the case with walls where stiction
helps to hold the steady state static configuration, we see that the mean slip
velocity closely follows the slope of the slip estimate. Without the walls,
objects do not pile up in a complex static structure but simply lie on the
ground, and therefore, the resulting slip velocities are significantly smaller.
The sudden drop in the slip velocity for $\sigma>10^{-3}$ is caused by the
sensitivity of the final state on the value of $\sigma$. As $\sigma$ increases,
so does the slip velocity bound $v_s$ and objects in the configuration without
walls can slowly drift into a configuration leading to more contacts. In
particular, boxes are more likely to slowly drift until one of their faces lies
flat on the ground, a configuration with zero slip once steady state is reached.

\begin{figure}[!h]
	\centering
    %trim={<left> <lower> <right> <upper>}
    \adjincludegraphics[width=0.49\columnwidth,trim={0 0 {0.05\width} 0},clip]{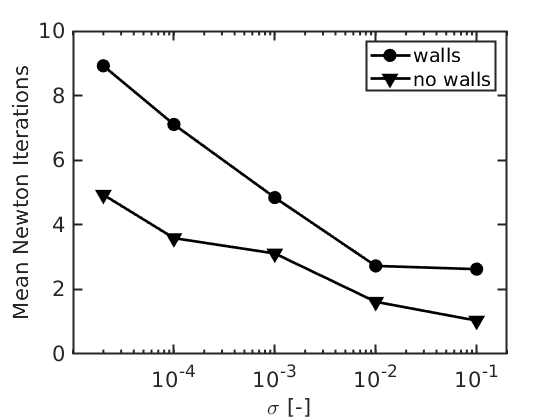}
    \adjincludegraphics[width=0.49\columnwidth,trim={0 0 {0.05\width} 0},clip]{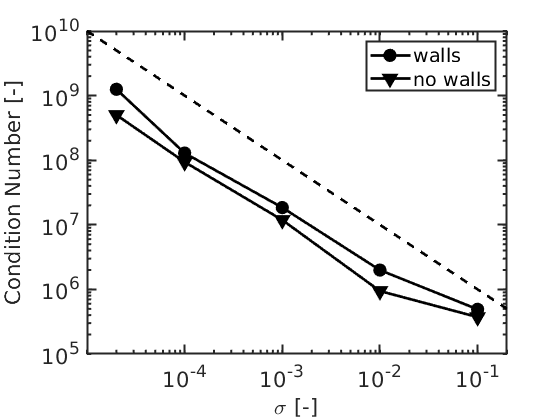}
	\caption{\label{fig:clutter_sigma} 
	Effect of the slip parameter on the mean Newton iterations per step (left)
	and mean condition number (right).}
\end{figure}

We conclude by examining the effect of $\sigma$ on the conditioning of the
system. Figure \ref{fig:clutter_sigma} shows the condition number of the Hessian
in the final configuration and the mean number of Newton iterations throughout
the simulation. We see that the condition number scales as $\sigma^{-1}$ while
the mean number of Newton iterations is roughly proportional to  $\ln(\sigma)$.
Our default choice $\sigma=10^{-3}$ is a good compromise between accurate
stiction, performance and conditioning.

\subsection{Slip Control}
\label{sec:slip_control}
While previous work on convex approximations model rigid contact
\cite{bib:anitescu2006,bib:mazhar2014} or use regularization as a means of
constraint stabilization \cite{bib:todorov2014}, our work is novel in that we
incorporate physical compliance. This allows us not only to model compliant
point contact, but also to incorporate sophisticated models of surfaces patches.
We incorporate the pressure field model \cite{bib:elandt2019pressure}
implemented as part of Drake's \cite{bib:drake} \emph{hydroelastic contact}
model. We use the discrete approximation introduced in
\cite{bib:masterjohn2021discrete} to approximate each face of the contact
surface as a compliant contact point at its centroid.

\begin{figure}[!h]
	\centering
	\includegraphics[width=0.8\columnwidth]{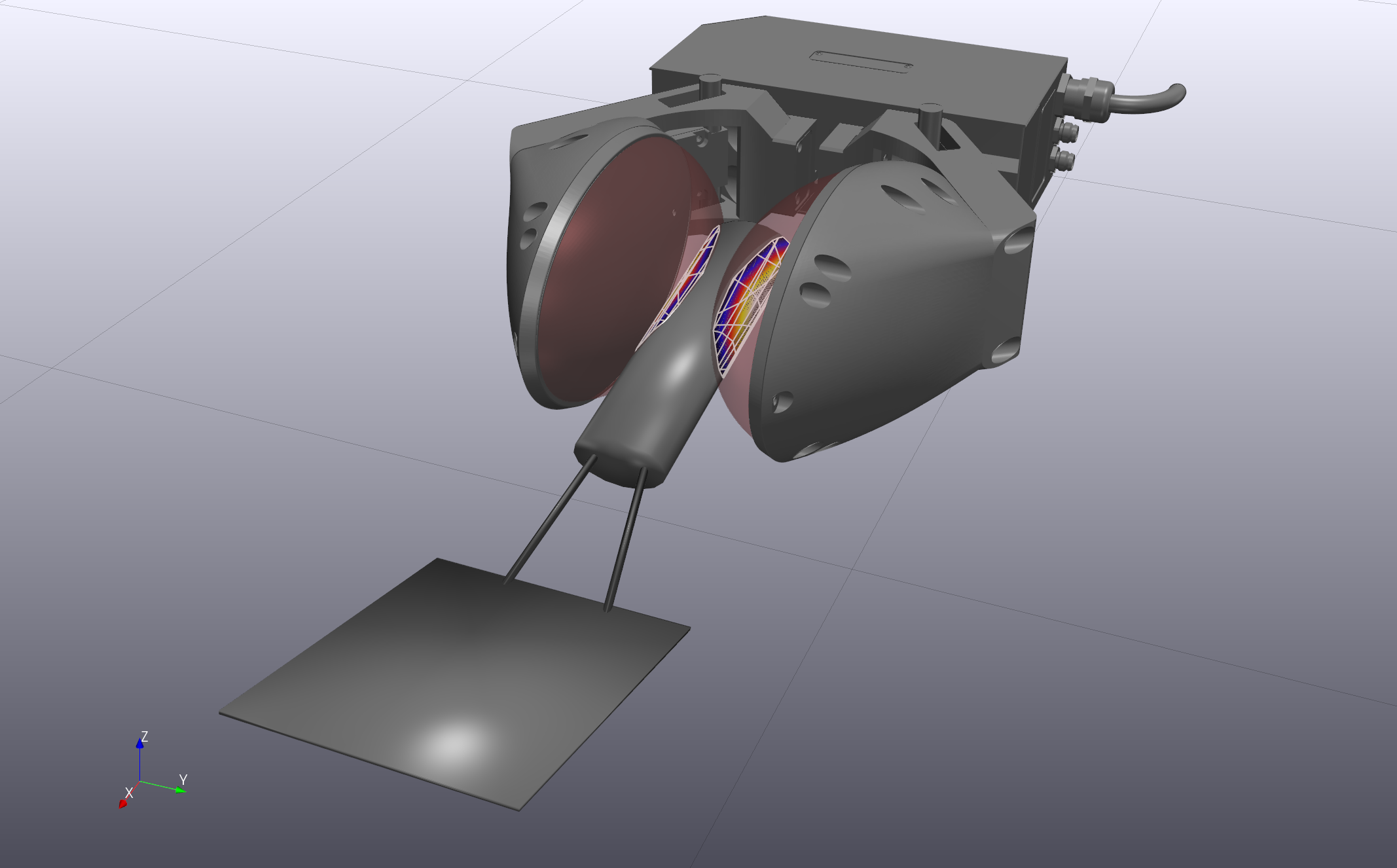}
	\caption{\label{fig:slip_control_frame} 
	Highly compliant \emph{Soft-bubble} gripper \cite{bib:kuppuswamy2020soft}
	holding a spatula. Unlike traditional point contact approaches, the
	hydroelastic contact model provides rich contact information and captures
	area-dependent phenomena such as the net-torque to hold the spatula. Contact
	patches are colored by contact pressure.}
\end{figure}

\begin{figure}[!h]
	\centering
    %trim={<left> <lower> <right> <upper>}
    \adjincludegraphics[width=0.49\columnwidth,trim={0 0 {0.05\width} 0},clip]{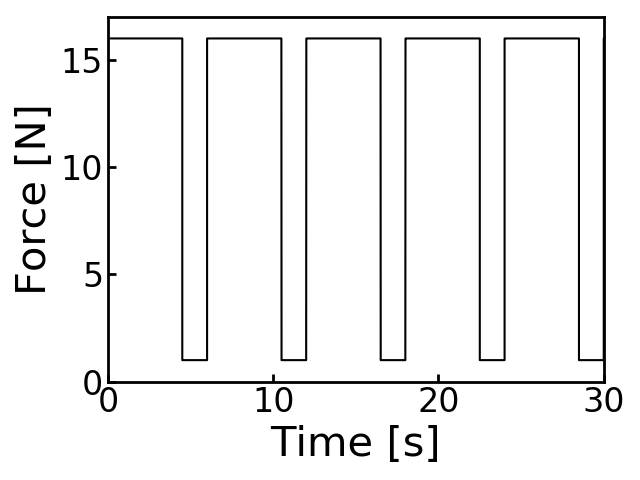}
    \adjincludegraphics[width=0.49\columnwidth,trim={0 0 {0.05\width} 0},clip]{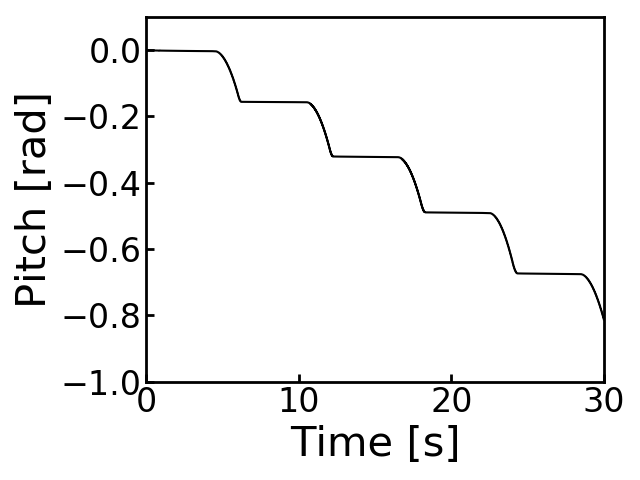}
	\caption{\label{fig:slip_control_history} 
	Grip force command (left) and spatula pitch angle (right) as a function of
	time.}
\end{figure}

To demonstrate this capability, we reproduce the test in
\cite{bib:masterjohn2021discrete} that models a \emph{Soft-bubble} gripper
\cite{bib:kuppuswamy2020soft}; a parallel jaw WSG 50 Schunk gripper outfitted
with air filled compliant surfaces (Fig. \ref{fig:slip_control_frame}). The aforementioned gripper is simulated
anchored to the world holding a spatula by the handle horizontally. We use
$\delta t=5\times 10^{-3}\text{ s}$. The grasp force is commanded to vary
between 1 N and 16 N with square wave having a 6 second period and a 75\% duty
cycle (see Fig.~\ref{fig:slip_control_history}, left). \reviewquestion{R8-Q3}{This
results in a periodic transition from a secure grip to a loose grip allowing the
spatula to pitch in a controlled manner within grasp (see
Fig.~\ref{fig:slip_control_history} and the accompanying video). These contact
mode transitions are resolved by our model.} We observe that stiction during the
secure grip is properly resolved with the tight bounds for the slip due to
regularization discussed in Section \ref{sec:conditioning}.
While this case only has 8 degrees of freedom, it generates about 60 contact
constraints during the slip phase and about 160 contact constraints during the
stiction phase.

We compare the performance of SAP against both Gurobi and Geodesic IPM, see
Section \ref{sec:about_solvers}. For SAP we use a relative tolerance
$\varepsilon_r=10^{-3}$, see Section \ref{sec:stopping_criteria}. For Gurobi we
set its tolerance parameter \verb+BarQCPConvTol+ to $10^{-8}$. For Geodesic IPM,
we set its complementary slackness tolerance to $10^{-6}$; larger values lead to
failure for this task. SAP bounds the momentum error, exhibiting a maximum value
of of $9.99\times 10^{-2}\,\%$. Even with such a tight tolerance, Gurobi
exhibits $2.6\,\%$ maximum error. Geodesic IPM's errors are significantly
smaller, below $2\times 10^{-4}\,\%$. However its robustness is very sensitive
to the specified tolerance.

Even though SAP's solutions are significantly more accurate than those from
Gurobi, it performs 92 times faster than Gurobi. SAP is 20 times faster than
Geodesic IPM and significantly more robust to solver tolerances. In terms of
solver iterations, SAP only performs 0.62 iterations on average, showcasing how
effectively it warm-starts. Geodesic IPM performs 5.6 iterations per step on
average and Gurobi performs 10.1 iterations on average.

\subsection{Dual Arm Manipulation}
\label{sec:dual_arm}
We demonstrate the effectiveness of our approach with the simulation of a
complex manipulation task. In this scenario, two Kuka IIWA arms (7 DOFs each)
are outfitted with anthropomorphic Allegro hands (16 DOFs each) (Fig.
\ref{fig:dual_arm_frames}). In front of the robot, a table has a jar (with a lid,
12 DOFs) full of 16 marbles of $50$~gr each (96 DOFs) and a bowl (6 DOFs),
completing the model with a total of 160 DOFs. Contact between the jar and the
lid is modeled using Drake's hydroelastic model \cite{bib:elandt2019pressure,
bib:masterjohn2021discrete} (see Section \ref{sec:slip_control}), while point
contact is used for all other interactions. The time step is set to $\delta
t=5\times 10^{-3}$~s.

The arms' controllers track a prescribed sequence of Cartesian end-effector
keyframe poses, while the hands' controllers track prescribed \emph{open/close}
configurations. We use force feedback to gauge successful grasps and to know
when the jar makes contact with the table. The robot is commanded to open the
jar, pour its contents into the bowl, close the lid and put the empty jar back
in place (keyframes in Fig. \ref{fig:dual_arm_frames} and the accompanying
video). 

\begin{figure}[!h]
	\centering
    \includegraphics[width=0.9\columnwidth]{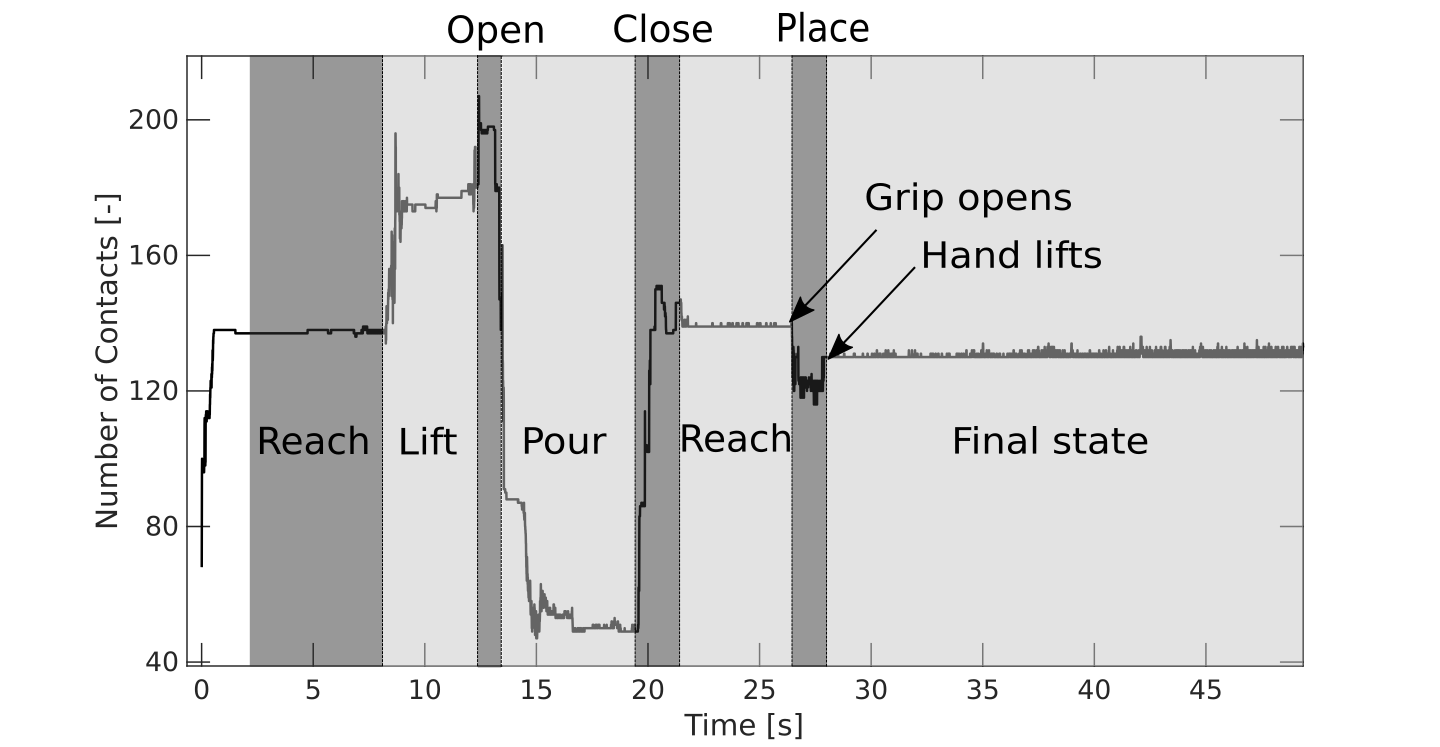}
    \caption{\label{fig:dual_arm_contacts} Number of contact constraints as a
    function time. Important events during the task are highlighted.}
\end{figure}

This particular task generates hundreds of contact constraints, as shown in Fig.
\ref{fig:dual_arm_contacts} which also labels important events during the task.
\reviewquestion{R8-Q3}{We remark that our framework predicts contact mode
switching as a result of the computation. For instance, the lid initially
covering the jar is held by stiction and it transitions to sliding as the robot
pulls it out.}

\begin{figure}[!h]
	\centering
    \includegraphics[width=0.7\columnwidth]{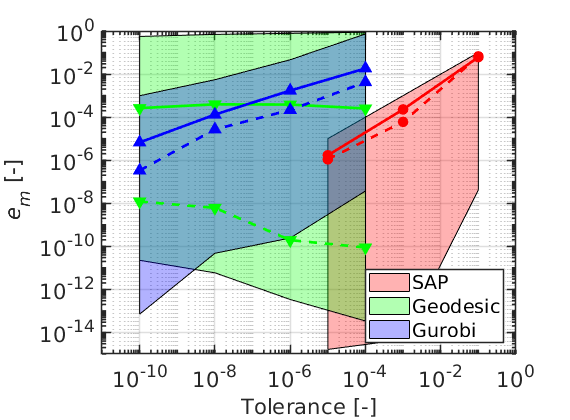}
    \caption{\label{fig:dual_arm_momentum} Dimensionless momentum error, defined
    in Eq. \eqref{eq:momentum_error}. Mean (solid) and median (dashed) errors
    along with minimum and maximum errors (shaded areas) over the entire
    simulation.}
\end{figure}

\begin{figure}[!h]
	\centering
    \includegraphics[width=0.7\columnwidth]{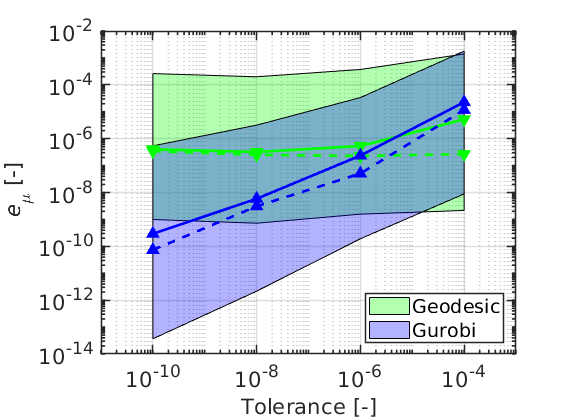}
    \caption{\label{fig:dual_arm_slackness} Dimensionless complementarity
    slackness error, defined in Eq. \eqref{eq:slackness_error}. Mean (solid) and
    median (dashed) errors along with minimum and maximum errors (shaded areas)
    over the entire simulation.}
\end{figure}

To assess accuracy, we evaluate the dimensionless momentum and complementarity
slackness errors defined in Eqs. \eqref{eq:momentum_error} and
\eqref{eq:slackness_error} respectively. We perform the simulation of the same
task several times using different solver tolerances. The results of these runs
are shown in Figures \ref{fig:dual_arm_momentum} and
\ref{fig:dual_arm_slackness}. Even though each solver uses a different tolerance
parameter, it is still useful to place these tolerances in the same horizontal
axis. The maximum tolerance we use for each solver corresponds to the largest
value that can be used to complete the task successfully. For Gurobi and
Geodesic IPM, smaller values of the tolerance parameter make the simulation
impractically slow. SAP on the other hand cannot achieve errors below $10^{-6}$
for this case due to round-off errors. Figures \ref{fig:dual_arm_momentum} and
\ref{fig:dual_arm_slackness} show both mean and median of the errors over the
entire simulation to show errors do not follow a symmetric distribution. More
interesting however are the minimum and maximum errors, shown as shaded areas.
SAP guarantees that momentum errors are below the specified tolerance, given
this is precisely its stopping criteria in Eq. \eqref{eq:stopping_criteria}. We
see however that it is difficult to correlate the expected error to solver
tolerance when using Gurobi or Geodesic IPM. In practice, we consistently
observe that the robot does not complete the task successfully when
momentum errors are larger than about $10\%$, regardless of the solver.
Therefore, we find that being able to specify a tolerance for the momentum
error directly is immensely useful. Given that SAP satisfies the complementarity
slackness condition exactly, complementarity slackness error for SAP is not
included in Fig. \ref{fig:dual_arm_slackness}.

\begin{figure}[!h]
	\centering
    \includegraphics[width=0.7\columnwidth]{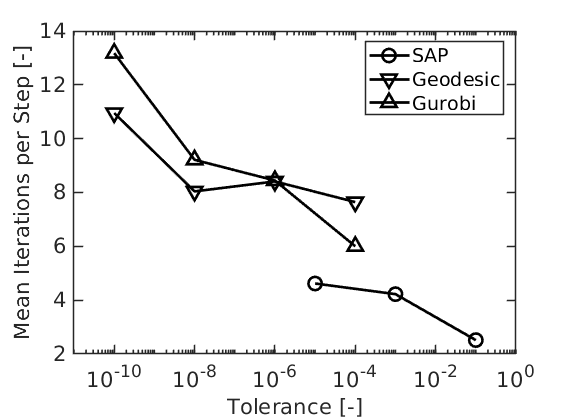}
    \caption{\label{fig:dual_arm_iterations} Mean number of iterations per
    time-step for the dual arm simulation.}
\end{figure}

Figure \ref{fig:dual_arm_iterations} shows the mean number of iterations per
time-step for each solver. We see that the number of iterations needed by the
SAP solver is consistently below the other two solvers given how effectively SAP
warm-starts from the previous time-step solution.

To make a fair comparison among solvers, from Fig. \ref{fig:dual_arm_momentum}
we choose tolerances for each solver that result in similar values of the mean
momentum error. For Gurobi, we set its tolerance parameter \verb+BarQCPConvTol+
to $10^{-8}$. For Geodesic IPM, we set its complementary slackness tolerance to
$10^{-6}$. For SAP, we set its relative tolerance to $10^{-3}$. Notice this is
not entirely fair to SAP, given that SAP does guarantee the maximum momentum
error to be below $10^{-3}$, while this is not true for the other two solvers.
Still, SAP is 7.4 faster than Gurobi and 2.2 faster than Geodesic IPM. In terms
of iterations, SAP performs 4 iterations on average while Geodesic IPM performs
8.3 iterations on average. This shows that since both solvers use exactly the
same sparse algebra, the performance gains with SAP are entirely due to its
ability to warm-start effectively rather than to differences in the
implementation. The general purpose solver Gurobi on the other hand performs
10.1 iterations on average.

In summary, the simulation of this complex robotic task demonstrates how
accuracy translates directly to robustness. We observe how the maximum momentum
error defined in Eq. \eqref{eq:momentum_error} is a good proxy for robustness in
simulation; simulations with errors larger than about $10\%$ could not complete
the task successfully. In this regard, SAP provides a certificate of accuracy
that proves useful in practice.

\section{Variations and Extensions}
\label{sec:variations_and_extensions}

The method presented in this paper can be extended in several ways:

\textbf{Expand the family of constraints:} No doubt contact constraints are the
most challenging. However, our method can be extended to include bilateral
constraints, PD controllers with force limits and even joint dry friction
\cite{bib:todorov2014}.

\textbf{Branch induced sparsity:} In this work we exploit sparsity only at the
tree level. However, branch sparsity can lead to additional performance.
Consider for instance a standing humanoid robot with a floating hip. Since arms
and the upper torso are not in contact with the ground, they can be eliminated
from the computation. Additional performance gains could be attained using
specialized algebra for multibody dynamics \cite{bib:carpentier2021}.

\textbf{Parallelization:} This work focuses on accuracy, robustness, and
convergence properties of the algorithm executed in a single thread. The sparse
algebra can be parallelized and, in particular, disjoint \emph{islands} of
bodies can be solved separately in different threads.

\reviewquestion{R4-Q3}{\textbf{Deformable FEM models:} Using the SAP solver for the modeling of deformable objects with
contact and friction is the topic of current research efforts by the authors.
FEM models lead to state dependent stiffness \eqref{eq:stiffness_matrix} and
damping \eqref{eq:dissipation_matrix} matrices with a complex structure that
requires specialized handling of sparsity. Moreover, modeling assumptions must
be carefully analyzed in order to ensure the positive definiteness of these
matrices used in our convex approximation of contact.}

\textbf{Differentiation:} Since forces are a continuous function of state, the
model is well suited for applications requiring gradients such as trajectory
optimization, machine learning, parameter estimation, and control.
Factorizations computed during forward dynamics can be reused when computing
gradients for a performant implementation.

\section{Limitations}
\label{sec:limitations}

All models are approximations of reality, while numerical methods can only
approximate our models. We list here the limitations we identify for our
method.

\textbf{Convex Approximation:} The convex approximation amounts to a
\emph{gliding effect} during sliding at a distance $\phi\sim \delta
t\mu\Vert\vf{v}_t\Vert$. Regularization leads to a model of
regularized friction, Eq. \eqref{eq:regularized_friction}. Details are provided
in Section \ref{sec:contact_modeling_parameters}.

\textbf{Stiffness and Dissipation:} \reviewquestion{R4-Q3}{Our method requires
stiffness $\mf{K}$ and damping $\mf{D}$ matrices to be SPD or SPD
approximations (see Section \ref{sec:two_stage_scheme}). 
For joint level spring-dampers, the exact  $\mf{K}$ and $\mf{D}$ can be used, but for
other forces such as those arising from a spatial arrangement of springs, an SPD approximation
must be made as $\mf{K}$ might not be SPD in certain configurations.}

\textbf{Linear Approximations:} \reviewquestion{AE/R1-Q1/R1-Q2}{Algorithm
\ref{alg:sap_time_stepping} evaluates the SPD gradient $\mf{A}$ once at
$\mf{v}^*$ at each time step. In other words, our method replaces the original
balance of momentum \eqref{eq:scheme_momentum} with its linear approximation
\eqref{eq:momentum_linearized}. This is exact for many important cases and
accurate to second-order in the general case. See Section
\ref{sec:unconstrained_convex_formulation} for details.}

\textbf{Delassus Operator Estimation:} \reviewquestion{R4-Q6}{In Section
\ref{sec:conditioning} we use a diagonal approximation of the Delassus operator
to estimate stiffness in the \emph{near-rigid} regime. Corner cases exist.
Consider a pile of books. While the inertia of a contact at the bottom of the
pile is estimated solely on the mass of one book, this contact is supporting the
weight of the entire stack. Stiffness is underestimated and user intervention is
needed to set proper parameters.}

\textbf{Scalability:} \reviewquestion{R4-Q7}{We see no reason SAP with the
direct supernodal algebra (Section \ref{sec:problem_sparsity}) could not scale
to thousands of bodies if there is structured sparsity. However, scalability needs to be studied further, along
with the usage of iterative solvers such as Conjugate Gradient (CG), widely
used in optimization.}

\section{Conclusion}
\label{sec:future_directions}

We presented a novel unconstrained convex formulation of compliant contact. In
this formulation constraints are eliminated using analytic formulae that we
developed. Our scheme incorporates the midpoint rule into a two-stage scheme,
with demonstrated second order accuracy. We rigorously characterized our
numerical approximations and the artifacts introduced by the convex
approximation of contact. We reported limitations of our method and discussed
extensions and areas of further research.

We showed that regularization maps to physical compliance, allowing us to
eliminate algorithmic parameters and to incorporate complex models of continuous
contact patches. Moreover, we studied the trade off between regularization and
numerical conditioning for the simulation of \emph{near-rigid} bodies and
the accurate resolution of stiction.

We presented SAP, a robust and performant solver that warm-starts very
effectively in practice. SAP globally converges at least at a linear-rate and
exhibits quadratic convergence when additional smoothness conditions are
satisfied. SAP can be up to 50 times faster than Gurobi in small problems with
up to a dozen objects and up to 10 times faster in medium sized problems with
about 100 objects. Even though SAP uses the supernodal algebra implemented for
Geodesic IPM, it performs at least two times faster due to its effective
warm-starts from the previous time-step solution. Moreover, SAP is significantly
more robust in practice given that it guarantees a hard bound on the error in
momentum, effectively providing a certificate of accuracy.

We have incorporated SAP into the open source robotics toolkit Drake
\cite{bib:drake}, and hope that the simulation and robotics communities can
benefit from our contribution.

% Appendices:
\appendices
\section{Proof of Proposition \ref{prop:gradient_of_m_approximation}}
\label{app:gradient_of_m_approximation}
The Taylor expansion of $\mf{m}(\mf{v})$ at $\mf{v}=\mf{v}^*$ reads
\begin{align}
	\mf{m}(\mf{v}) &= \mf{m}^* + \frac{\partial \mf{m}}{\partial \mf{v}} (\mf{v}-\mf{v}^*) +
	\mathcal{O}_m(\Vert\mf{v}-\mf{v}^*\Vert^2)\nonumber\\
	&=\frac{\partial \mf{m}}{\partial \mf{v}}(\mf{v}-\mf{v}^*) +
	\mathcal{O}_m(\Vert\mf{v}-\mf{v}^*\Vert^2),
	\label{eq:m_taylor_expansion}
\end{align}
where we use the fact that by definition $\mf{m}^*=\mf{m}(\mf{v}^*)=\mf{0}$. All
derivatives are evaluated at $\mf{v} = \mf{v}^*$ unless otherwise noted. We
first evaluate the Jacobian of the mass matrix term in Eq.
(\ref{eq:m_definition})
\begin{align*}
	\frac{\partial \left( \mf{M}(\mf{q}^{\theta}(\mf{v}))(\mf{v}-\mf{v}_0) \right)}{\partial \mf{v}}
	= \mf{M}(\mf{q}^{\theta}(\mf{v}^*)) + \mf{E},
\end{align*}
where we defined
\begin{align*}
	\mf{E} = \frac{\partial \mf{M}(\mf{q}^{\theta})}{\partial\mf{v}} (\mf{v}^*-\mf{v}_0).
\end{align*}
Note that by combining Eqs. (\ref{eq:theta_method}) and (\ref{eq:scheme_q_update}), the
mid-step configuration $\mf{q}^{\theta}$ can be written as
\begin{align*}
	\mf{q}^{\theta}(\mf{v}) &= \mf{q}_0 + \delta t \theta \dot{\mf{q}}^{\theta_{vq}} \\
	                          &= \mf{q}_0 + \delta t \theta \mf{N}(\mf{q}^{\theta})\mf{v}^{\theta_{vq}}(\mf{v}).
\end{align*}
Hence by the chain rule, $\mf{E}$ can be further calculated as
\begin{align*}
	\mf{E} = \delta t\theta\frac{\partial \mf{M}(\mf{q}^{\theta}) }{\partial\mf{q}}
             \frac{\partial\dot{\mf{q}}^{\theta_{vq}}}{\partial\mf{v}}
			 (\mf{v}^*-\mf{v}_0).
\end{align*}
Notice that 
\begin{align*}
		\Vert\mf{E}\Vert 
		&\le \delta t\theta \left\| \frac{\partial\mf{M}(\mf{q}^{\theta})}{\partial\mf{q}}  \right\|
			\left\| \frac{\partial\dot{\mf{q}}^{\theta_{vq}}}{\partial\mf{v}}  \right\|
		    \left\| \mf{v}^*-\mf{v}_0 \right\| \\
		&= \mathcal{O}(\delta t^2),
\end{align*}
since $\Vert\mf{v}^*-\mf{v}_0\Vert = \mathcal{O}(\delta t)$.

We proceed similarly to expand the Jacobian of
$\mf{F}_1(\mf{v})=\mf{F}_1(\mf{q}^{\theta}(\mf{v}), \mf{v}^{\theta}(\mf{v}))$
as
\begin{align*}
	\frac{\partial\mf{F}_1}{\partial \mf{v}} = -\delta t\,\theta\theta_{vq}\mf{K}(\mf{q}^{\theta},
	\mf{v}^{\theta})-\theta\mf{D}(\mf{q}^{\theta}, \mf{v}^{\theta}),
\end{align*}
with $\mf{K}$ and $\mf{D}$ the stiffness and damping matrices defined by Eqs.
(\ref{eq:stiffness_matrix})-(\ref{eq:dissipation_matrix}).

We can now write the Jacobian of $\mf{m}(\mf{v})$ in Eq.
(\ref{eq:m_taylor_expansion}) as
\begin{align*}
	\frac{\partial \mf{m}}{\partial \mf{v}} = \mf{A} + \mf{E} - \delta t\frac{\partial \mf{F}_2}{\partial \mf{v}},
\end{align*}
where we defined
\begin{align*}
	\mf{A}=\mf{M}+ \delta t^2\theta\theta_{qv}\mf{K}+\delta t\theta\mf{D}.
\end{align*}
With these definitions the Taylor expansion in Eq. (\ref{eq:m_taylor_expansion})
becomes
\begin{align*}
	\frac{\partial\mf{m}}{\partial\mf{v}}(\mf{v}-\mf{v}^*) &= \mf{A}(\mf{v}-\mf{v}^*) + \mf{E}(\mf{v}-\mf{v}^*) \\
	&- \delta t\frac{\partial\mf{F}_2}{\partial\mf{v}}(\mf{v}-\mf{v}^*) + \mathcal{O}_m(\Vert\mf{v}-\mf{v}^*\Vert^2).
\end{align*}

Since contact is compliant, forces are finite within the finite interval $\delta
t$ and therefore $\Vert\mf{v}-\mf{v}^*\Vert=\mathcal{O}(\delta t)$. Thus
\begin{align*}
	\mf{E}(\mf{v}-\mf{v}^*)=\mathcal{O}_E(\delta t^3), \\
    \delta t\frac{\partial \mf{F}_2}{\partial \mf{v}}(\mf{v}-\mf{v}^*)=\mathcal{O}_{F_2}(\delta t^2), \\ 
    \mathcal{O}_m(\Vert\mf{v}-\mf{v}^*\Vert^2) = \mathcal{O}_m(\delta t^2).
\end{align*}
Therefore, the positive definite linearization
\begin{align*}
	\mf{A}(\mf{v}-\mf{v}^*) + \mathcal{O}_E(\delta t^3) + \mathcal{O}_{F_2}(\delta t^2) +
	\mathcal{O}_m(\delta t^2) = \mf{J}^T\mf{\bgamma},
\end{align*}
agrees with the original momentum balance in Eq. (\ref{eq:scheme_momentum}) to second
order.

Finally, notice that $\mf{A}$ is a linear combination of positive definite
matrices with non-negative scalars in the linear combination, and therefore
$\mf{A}\succ 0$.\hfill\IEEEQED

\section{Proof of Theorem \ref{th:unconstrained_formulation_equivalance}}
\label{app:unconstrained_formulation_equivalance}
Before proving this theorem, we need the following result.
\begin{lemma}
    The conic constraint $\mf{g}(\mf{v}, \bsigma)\in\mathcal{F}^*$ is satisfied
    if $\bsigma$ is given by $P_\mathcal{F}(\mf{y(\mf{v})})$.
    \label{lemma:conic_constraints_are_satisfied_analytically}
\end{lemma}
\begin{IEEEproof}
    Since $\bsigma$ is the projection of $\mf{y}(\mf{v})$ to the cone
    $\mathcal{F}$ with the $\mf{R}$ norm, by Moreau's decomposition theorem, we
    know that $\mf{y}(\mf{v}) - \bsigma$ is in the polar cone of $\mathcal{F}$
    with the $\mf{R}$ norm. That is, $\langle \mf{y}(\mf{v}) - \bsigma, \mf{x} \rangle_\mf{R} \le 0$ for all $\mf{x} \in \mathcal{F}$, with the inner product
    $\langle\mf{v},\mf{w}\rangle_\mf{R}=\mf{v}^T\mf{R}\mf{w}$. Reorganizing
    terms, we get
    \begin{align*}
        \mf{x}^T \mf{R}(\mf{y}(\mf{v}) - \bsigma) &\le 0, \\
        \mf{x}^T (-\mf{R}\bsigma - \mf{v}_c + \hat{\mf{v}}_c) &\le 0, \\
        \langle \mf{x}, -\mf{R}\bsigma - \mf{v}_c + \hat{\mf{v}}_c \rangle &\le 0,
    \end{align*}
    for all $\mf{x} \in \mathcal{F}$. Therefore, it follows that
    $-\mf{g}=-(\mf{v}_c - \hat{\mf{v}}_c + \mf{R}\bsigma)$ is in the polar cone
    of $\mathcal{F}$ and thus $\mf{g}$ is in the dual cone of $\mathcal{F}$.
\end{IEEEproof}

The optimality condition for the unconstrained formulation in
\eqref{eq:primal_unconstrained} is $\nabla\ell_p(\mf{v})=\mf{0}$. It is shown in
Appendix \ref{app:gradients_derivation} that
\begin{equation*}
    \nabla\ell_p(\mf{v})=\mf{A}(\mf{v}-\mf{v}^*) - \mf{J}^T\bgamma(\mf{v}),
\end{equation*}
with impulses given
by $\bgamma(\mf{v})=P_\mathcal{F}(\mf{y}(\mf{v}))$, the dual optimal. Therefore,
$\nabla\ell_p(\mf{v})=\mf{0}$ implies \eqref{eq:momentum_optimality}, the first
optimality condition for \eqref{eq:primal_regularized}.
   
The analytical inverse dynamics solution shows that $\bgamma =
P_\mathcal{F}(\mf{y(\mf{v})})$ with the primal optimal $\mf{v}$. Hence, choosing
$\bsigma = P_\mathcal{F}(\mf{y(\mf{v})})$ with the primal optimal $\mf{v}$
satisfies \eqref{eq:sigma_equal_gamma}, the second optimality condition for
\eqref{eq:primal_regularized}.

Finally, by Lemma \ref{lemma:conic_constraints_are_satisfied_analytically}, the
cone constraint $\mf{g}(\mf{v}, \bsigma)\in\mathcal{F}^*$ is satisfied.
\hfill\IEEEQED

\section{Analytical Inverse Dynamics}
\label{app:analytical_inverse_dynamics_derivations}
We perform the projection in Eq. \eqref{eq:y_projection} for a regularization of
the form $\vf{R} = \text{diag}([R_t, R_t, R_n])$. For simplicity, we drop
contact subindex $i$. We make the change of variables
$\tilde{\bgamma}=\vf{R}^{1/2}\bgamma$ and $\tilde{\vf{y}}=\vf{R}^{1/2}\vf{y}$
\cite{bib:todorov2014}, and observe that $\tilde{\bgamma}$ is the Euclidian
projection of $\tilde{\vf{y}}$ onto cone $\tilde{\mathcal{F}}$ with coefficient
$\tilde \mu =\mu\,(R_t/R_n)^{1/2}$. We conclude that
\begin{eqnarray*}
  P_\mathcal{F}(\vf{y})=\vf{R}^{-1/2} P_{\tilde{\mathcal{F}}}(\tilde{\vf{y}}),
\end{eqnarray*}

\begin{figure}[!h]
    \centering
    %\vspace{6pt}
    \includegraphics[width=0.5\columnwidth]{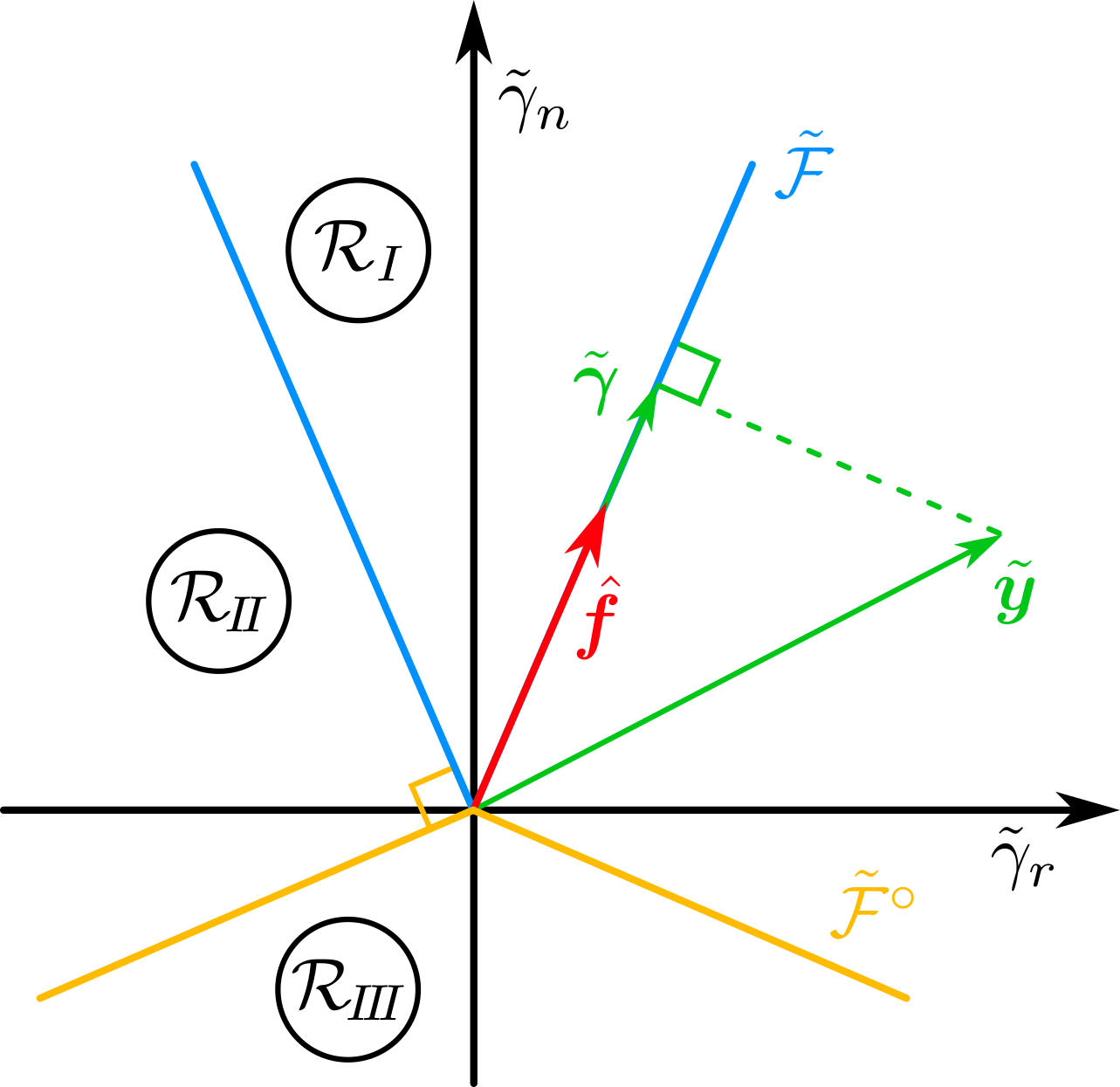}
    \caption{Geometry of the projection and regions in the
    $\tilde{\vf{y}}$ space.}
    \label{fig:cone_regions}
\end{figure}

We partition $\mathbb{R}^3$ into three regions, see Fig.~
\ref{fig:cone_regions}: closed cone $\tilde{\mathcal{F}}$, denoted with
$\mathcal{R}_I$, the interior of the polar $\tilde{\mathcal{F}}^\circ$, denoted
with $\mathcal{R}_{III}$, and the remaining area, which we denote with
$\mathcal{R}_{II}$. For $\tilde{\vf{y}}\in\mathcal{R}_I$ we simply have that
$P_{\tilde{\mathcal{F}}}(\tilde{\vf{y}}) = \tilde{\vf{y}}$. When
$\tilde{\vf{y}}\in\mathcal{R}_{III}$, $P_{\tilde{\mathcal{F}}}(\tilde{\vf{y}}) =
\vf{0}$. Finally, when $\tilde{\vf{y}}\in\mathcal{R}_{II}$, we evaluate
$P_{\tilde{\mathcal{F}}}(\tilde{\vf{y}})$ via Euclidean projection onto the
boundary of $\tilde{\mathcal{F}}$, which admits a simple formula. We define
$\hat{\vf{f}}=[\tilde{\mu}\hat{\vf{t}}, 1]/\sqrt{1+\tilde{\mu}^2}$, the unit
vector along the wall of the cone shown in Fig.~\ref{fig:cone_regions}, with
$\hat{\vf{t}}=\tilde{\vf{y}}_t/\Vert\tilde{\vf{y}}_t\Vert=\vf{y}_t/\Vert\vf{y}_t\Vert$.
Then the projection is computed as
$\tilde{\bgamma}=(\tilde{\vf{y}}\cdot\hat{\vf{f}})\hat{\vf{f}}$. After some
algebraic manipulation we have that $P_{\tilde{\mathcal{F}}}(\tilde{\vf{y}}) =
[\tilde{\bgamma}_t, \tilde{\gamma}_n]$ with
\begin{align*}
  \tilde{\bgamma}_t       &= \tilde{\mu}\tilde{\gamma}_n\hat{\vf{t}},\\
        \tilde{\gamma}_n  &= \frac{1}{1+\tilde{\mu}^2}\left(\tilde{y}_n +
	\tilde{\mu}\tilde{y}_r\right),
\end{align*}
where $\tilde{y}_r=\Vert\tilde{\vf{y}}_t\Vert$. Note that this formula is
well-defined on $\mathcal{R}_{II}$, since $\vf{y}_t = \vf{0}$ only if $\vf{y}$ is in regions $\mathcal{R}_{I}$ or $\mathcal{R}_{III}$.

Finally, we apply the inverse transformation
$\bgamma=\mf{R}^{-1/2}P_{\tilde{\mathcal{F}}}(\tilde{\vf{y}})$ and after some
algebraic manipulation we recover the projection $\bgamma =
P_\mathcal{F}(\vf{y})$ in Eq. \eqref{eq:analytical_y_projection}.

\section{Computation of the Gradients}
\label{app:gradients_derivation}
We will start by taking derivatives of the regularizer term $\ell_R$. First we
notice that we can write this term as
\begin{equation*}
	\ell_R = \frac{1}{2}\Vert\bgamma\Vert_{R}^2 = 
	\sum_i \ell_{R_i},
\end{equation*}
with $\ell_{R_i}=1/2\Vert\bgamma_i\Vert_{R_i}^2$. Since
$\nabla_{\vf{y}_j}\ell_{R_i}=\vf{0}$ for $i\neq j$, we only need to compute the
gradients of $\ell_{R_i}(\mf{y})$ with respect to the contact point variable
$\vf{y}_i\in\mathbb{R}^3$. Dropping contact subindex $i$ for simplicity, we
write the regularization as
\begin{eqnarray*}
	\ell_R = \frac{1}{2}\Vert\bgamma\Vert_R^2=\frac{1}{2}(R_t\Vert\bgamma_t\Vert^2+R_n\gamma_n^2).
\end{eqnarray*}

We use Eq.~(\ref{eq:analytical_y_projection}) to write the cost in terms of $\vf{y}$ as
\begin{align}
	&\ell_R(\vf{y}) = 
	\label{eq:ell_R_piecewise}\\	
&\begin{dcases}
	% Region I, stiction
	\frac{1}{2}(R_t y_r^2+R_n y_n^2) & \text{stiction, } y_r \le \mu y_n\\
	% Region II, sliding.
	\frac{R_n}{2(1+\tilde\mu^2)}\left(y_n + \hat\mu y_r\right)^2 & \text{sliding, } -\hat\mu y_r < y_n \leq \frac{y_r}{\mu}\\
	% Region II, no contact.
    \vf{0} & \text{no contact, } y_n < -\hat\mu y_r
\end{dcases}\nonumber	
\end{align}

\subsection{Gradients per Contact Point}
We use the following identities to simplify expressions
\begin{equation*}
	\frac{\partial y_r}{\partial\vf{y}_t} = \hat{\vf{t}}\nonumber,
	\quad
	\frac{\partial \hat{\vf{t}}}{\partial\vf{y}_t} =
	\frac{\vf{P}^\perp(\hat{\vf{t}})}{y_r},
\end{equation*}
where the $2\times 2$ projection matrix is
\begin{equation*}
	\vf{P}^\perp(\hat{\vf{t}})=\vf{I}_2 - \vf{P}(\hat{\vf{t}}),
	\quad\text{with }
	\vf{P}(\hat{\vf{t}}) = \hat{\vf{t}}\otimes\hat{\vf{t}}\nonumber.
\end{equation*}

Taking the gradient of Eq. (\ref{eq:ell_R_piecewise}) results in
\begin{align}
	&\nabla_\vf{y}\ell_R(\vf{y}) = 
	\label{eq:gradient_ell_R_piecewise}\\
&\begin{dcases}
	%%%%%%%%%%%%%%%%%%%%
	% Region I, stiction
	\vf{R}\,\vf{y} & 
	% when,
	\text{stiction, } y_r \le \mu y_n\\
	%
	%%%%%%%%%%%%%%%%%%%%
	% Region II, sliding.
	\frac{1}{1+\tilde\mu^2}\hat{s}^\circ(\vf{y})\begin{bmatrix}
		\mu R_t\hat{\vf{t}}\\
		R_n\\
	\end{bmatrix} &
	% when,
	\text{sliding, } -\hat\mu y_r < y_n \leq \frac{y_r}{\mu}\\
	% Region II, no contact.
    \vf{0} & \text{no contact, } y_n < -\hat\mu y_r
\end{dcases}\nonumber
\end{align}
with $\hat{s}^\circ(\vf{y}) = \hat{\mu}y_r+y_n$ positive in the sliding region. We note that $\nabla_\vf{y}\ell_R(\vf{y})$ is not differentiable at the
boundaries of $\mathcal{F}$ and $\mathcal{F}^\circ$. At points of
differentiability, the Hessian $\nabla_\vf{y}^2\ell_R(\vf{y})$ is computed by
taking derivatives of Eq. (\ref{eq:gradient_ell_R_piecewise})
\begin{align}
	&\nabla_\vf{y}^2\ell_R(\vf{y}) = 
	\label{eq:hessian_ell_R_piecewise}\\
&\begin{dcases}
	%%%%%%%%%%%%%%%%%%%%
	% Region I, stiction
	\vf{R} & 
	% when,
	\text{stiction,}\\
	%
	%%%%%%%%%%%%%%%%%%%%
	% Region II, sliding.	
	\frac{R_n}{1+\tilde\mu^2}
	\begin{bmatrix}
		% ∂²ℓ/∂yₜ²:
		\hat{\mu}\left(\hat{\mu}\vf{P}(\hat{\vf{t}})+\hat{s}^\circ(\vf{y})\vf{P}^\perp(\hat{\vf{t}})/y_r\right) & 
		% ∂²ℓ/∂yₙ∂yₜ:
		\hat{\mu}\vf{t}\\
		% ∂²ℓ/∂yₜ∂yₙ:
		\hat{\mu}\vf{t}^T & 
		% ∂²ℓ/∂yₙ²:
		1\\
	\end{bmatrix} &
	% when,
	\text{sliding,}\\
	% Region II, no contact.
    \vf{0} & \text{no contact.}
\end{dcases}\nonumber
\end{align}

Clearly in the stiction region we have $\nabla_\vf{y}^2\ell_R(\vf{y})\succ 0$.
Since in the stiction region we have $\hat{s}^\circ(\vf{y})>0$, the linear
combination of $\vf{P}(\hat{\vf{t}})$ and $\vf{P}(\hat{\vf{t}})^\perp$ in Eq.
(\ref{eq:hessian_ell_R_piecewise}) is PSD (since both projection matrices are
PSD). Therefore $\nabla_\vf{y}^2\ell_R(\vf{y})\succeq 0$.

\subsection{Gradients with Respect to Velocities}
Recall we use bold italics for vectors in $\mathbb{R}^3$ and non-italics bold
for their stacked counterpart. With $\mf{y}=-\mf{R}^{-1}(\mf{J}\mf{v} -
\hat{\mf{v}}_c)$ we use the chain rule to compute the gradient in terms of
velocities 
\begin{equation}
	\nabla_\mf{v}\ell_R = -\mf{J}^T\mf{R}^{-1}\nabla_\mf{y}\ell_R,
	\label{eq:ell_velocity_gradient}
\end{equation}
which, using Eq. (\ref{eq:gradient_ell_R_piecewise}), can be shown to equal
\begin{equation}
	\nabla_\mf{v}\ell_R = -\mf{J}^T\bgamma,
	\label{eq:ell_velocity_gradient_simplified}
\end{equation}

At points of differentiability, we obtain the Hessian of the regularizer
$\ell_R(\mf{v})$ from the gradient of $\bgamma(\mf{v})$ in Eq.
\eqref{eq:ell_velocity_gradient_simplified}
\begin{align*}
	\nabla_\mf{v}^2\ell_R(\mf{v}) &= -\mf{J}^T \nabla_{\mf{v}_c}\bgamma\,\mf{J}\nonumber,\\
	\nabla_{\mf{v}_c}\bgamma &= -\nabla_\mf{y}\bgamma \mf{R}^{-1},
\end{align*}
where $\nabla_{\mf{v}_c}\!\bgamma$ is a block diagonal matrix where each
diagonal block is the $3\times 3$ matrix $\nabla_{\mf{v}_{c,i}}\!\bgamma_i$
for the $i\text{-th}$ contact. Alternatively, taking the gradient of Eq. \eqref{eq:ell_velocity_gradient} leads to the equivalent result
\begin{equation*}
	\nabla_\mf{v}^2\ell_R = \mf{J}^T\mf{R}^{-1}\nabla_\mf{y}^2\ell_R\mf{R}^{-1}\mf{J},
\end{equation*}
where we can verify indeed that
$-\nabla_{\mf{v}_c}\bgamma=\mf{R}^{-1}\nabla_\mf{y}^2\ell_R\mf{R}^{-1}$. Since
$\nabla_\mf{y}^2\ell_R\succeq 0$, it follows that
$-\nabla_{\mf{v}_c}\bgamma\succeq 0$.

We define $\vf{G}_i\in\mathbb{R}^{3\times 3}$ the matrix that evaluates to
$-\nabla_{\mf{v}_{c,i}}\!\bgamma_i$ within regions $\mathcal{R}_I$,
$\mathcal{R}_{II}$ and $\mathcal{R}_{III}$ where the projection is
differentiable. At the boundary of $\mathcal{F}$ we use the analytical
expression from $\mathcal{R}_I$. At the boundary of $\mathcal{F}^\circ$ we use
the analytical expression from $\mathcal{R}_{II}$. This extension fully
specifies $\vf{G}_i$ for all $\vf{y}_i\in\mathbb{R}^3$. Finally, we define the
${3n_c\times 3n_c}$ matrix $\mf{G}=\text{diag}(\vf{G}_i)\succeq 0$.

\subsection{Gradients of the Primal Cost}
With these results, we can now write the gradient $\nabla_\vf{v}\ell_p$ and weighting matrix $\mf{H}$ in Algorithm \ref{alg:sap}. For the gradient we
have
\begin{equation*}
	\nabla_\mf{v}\ell_p(\mf{v}) = \mf{A}(\mf{v}-\mf{v}^*) + \nabla_\mf{v}\ell_R.
\end{equation*}
which using Eq. (\ref{eq:ell_velocity_gradient_simplified}) can be written as
\begin{equation*}
	\nabla_\mf{v}\ell_p(\mf{v}) = \mf{A}(\mf{v}-\mf{v}^*) - \mf{J}^T\bgamma,
\end{equation*}
and since the unconstrained minimization seeks to satisfy the optimality
condition $\nabla_\mf{v}\ell_p=\mf{0}$, we recover the balance of momentum.

Finally, we define the weighting matrix $\mf{H}$ as
\begin{equation*}
	\mf{H} = \mf{A} + \mf{J}^T\mf{G}\,\mf{J},
\end{equation*}
which, given the definition of $\mf{G}$, returns the Hessian of $\ell_p(\vf{v})$
when the gradient is differentiable and extends the analytical expressions at
points of non-differentiability. Since $\mf{A}\succ 0$ and $\mf{G} \succeq 0$,
we have $\mf{H}\succ 0$.

\section{Convergence Analysis of SAP}
\label{app:sap_converge}
\newcommand{\coneName}{\mathcal{K}}
\newcommand{\dist}{d}
\newcommand{\cond}{\text{cond}}
\newcommand{\vx}{\mf{v}}
\newcommand{\fx}{\ell_p(\mf{\vx})}

\newcommand{\vy}{\mf{u}}
\newcommand{\fy}{\ell_p(\mf{\vy})}
\newcommand{\vd}{\mf{d}}

% As required by the IEEE template.
\renewcommand\qedsymbol{$\IEEEQED$}

Convergence of SAP is established
by first showing that the objective function 
$\ell_p(\mf{v}) = \frac{1}{2}\Vert\mf{v}-\mf{v}^*\Vert_{A}^2 + P_\mathcal{F}(\mf{y}(\mf{v}))\Vert_R^2$
is \emph{strongly convex}
and differentiable with \emph{Lipschitz continuous} gradients.  The
former property is inherited from the positive-definite quadratic term 
provided by the positive definite matrix $\mf{A}$ in Eq. \eqref{eq:primal_unconstrained}.  The latter is shown using differentiability of the squared-distance function
 and the Lipschitz continuity of its gradient map (Theorems 5.3-i 6.1-i of~\cite{bib:delfour2011shapes})
 combined with the identity
\[
  \dist^2_{\coneName^\circ}(x) = \|P_{\coneName}(x)\|_{\mf{R}}^2,
\] 
for any closed, convex cone $\coneName$. Here 
the distance and projection functions are with respect to the
norm $\|\cdot\|_\mf{R}$, while $\coneName^\circ$ denotes the polar
cone with respect to the corresponding inner-product $\mf{x}^T \mf{R} \mf{y}$.
\begin{lemma}\label{lem:PropertiesOfObj}
  The following statements hold.
  \begin{itemize}
    \item The function $\fx$ is strongly convex, i.e., there exists $\mu >0$
      such that 
      \[
        \fy \ge \fx + \nabla \fx(\vy-\vx) + \frac{\mu}{2} \|\vy-\vx\|^2
      \]
    \item The function $\fx$ is differentiable and has Lipschitz continuous gradients, i.e.,
      $\nabla \fx$ exists for all $\mf{v}$ and there exists $L \ge 0$ satisfying
      \[
      \|\nabla \fx - \nabla \fy\| \le L \|\vy - \vx\|
      \]
  \end{itemize}
  \begin{proof}

The objective $\fx$ is a function $f : \mathbb{R}^n \rightarrow \mathbb{R}$
of the following form
\[
  f(\vx) = \frac{1}{2}\dist_{\coneName}^2(\mf{Z}\vx + \mf{b}) + \vx^{T}\mf{W}\vx + \mf{q}^T \vx,
\]
where $\mf{Z} \in \mathbb{R}^{m \times n}$, $\mf{W} \in \mathbb{R}^{n \times n}$ is
symmetric and positive definite, and $\dist_{\coneName} : \mathbb{R}^{m} \rightarrow \mathbb{R}$ 
denotes the distance function of a closed, convex set $\coneName \subseteq \mathbb{R}^m$
as measured by some quadratic norm $\|\mf{x}\|_Q$, i.e.,
\[
  \dist_{\coneName}(\vx) = \inf \{ \|\vx-\mf{z}\|_Q : \mf{z} \in \coneName\}.
\]
    The sum of a strongly convex function with a convex function is strongly
    convex.  Since the squared distance function is convex, and the quadratic
    term $\vx^T \mf{W}\vx$ is strongly convex (given that $\mf{W} \succ 0$), the  first
    statement holds.

    The second statement follows trivially if we can show it holds 
    for the squared distance function. 
    Differentiability follows from Chapter 4 (Theorems 5.3-i 6.1-i) of~\cite{bib:delfour2011shapes},
    which shows that
  \[
    \nabla \dist_{\coneName}^2(\vx) = 2 \mf{Q} (\vx - P_{\coneName}(\vx)).
  \]
    That the gradient of $\dist_{\coneName}^2(\vx)$ is Lipschitz follows
    from Lipschitz continuity of projection maps onto closed, convex sets.
  \end{proof}
\end{lemma}
\noindent We remark that strong convexity implies the reverse
Lipschitz inequality $\|\nabla f(\vx) - \nabla f(\vy)\| \ge \mu \|\vx - \vy\|$,
which in turn means that the parameter $\mu$ and the Lipschitz constant $L$
satisfy $\mu \le L$.

Recall that SAP (Algorithm~\ref{alg:sap}) is a special case
of the following iterative method
for minimizing a function $f : \mathbb{R}^n \rightarrow
\mathbb{R}$ given some initial
point $\vx_0 \in \mathbb{R}^n$:
\begin{align}\label{eq:quasiNewtonIter}
  \begin{aligned}
    \vd_m &= -\mf{H}^{-1}(\vx_m) \nabla f(\vx_m), \\
     t_m &= \argmin_{t} f(\vx_m + t \vd_m),\\
    \vx_{m+1} &= \vx_m + t_m \vd_m,      
  \end{aligned}
\end{align}
where $\mf{H} : \mathbb{R}^n \rightarrow \mathbb{R}^{n \times n}$ is a function 
into the set of symmetric positive definite matrices, i.e.,  $\mf{H}(\vx) = \mf{H}(\vx)^T$ 
and $\mf{H}(\vx) \succ 0$ for all $\vx\in\mathbb{R}^n$.  
It is well known that gradient descent exhibits linear convergence to the global minimum when applied to a strongly convex
function with Lipschitz continuous gradient. Incorporating
a condition number bound  $\sigma$ for $\mf{H}(\vx)$  into standard gradient-descent analysis
will prove that the iterations~\eqref{eq:quasiNewtonIter}
also have linear convergence.
To show this, we let
$\cond(\mf{H}(\vx))$ denote the condition number of $\mf{H}(\vx)$
and $S(\vx_0)$ denote the sub-level set $\{ \vx \in \mathbb{R}^n: f(\vx) \le f(\vx_0) \}$.
\begin{lemma}\label{lem:GlobalConv}
  Let $f : \mathbb{R}^n \rightarrow \mathbb{R}$ be strongly convex and differentiable
  with Lipschitz-continuous gradients.
  Fix $\vx_0 \in\mathbb{R}^n$. If there exists $\sigma > 0$
  such that $\cond(\mf{H}(\vx)) \le \sigma$ for all  $\vx \in S(\vx_0)$, then
  the iterations~\eqref{eq:quasiNewtonIter}
  converge to the global minimum $\vx_*$ of $f(\vx)$ when initialized at $\vx_0$.
  Moreover,
  \[
    f(\vx_m)  - f(\vx_*)  \le (1-\frac{\mu}{\sigma^2 L})^m (f(\vx_0) - f(\vx_*))
  \]
  for all iterations $m$, where $\mu$ is the strong-convexity parameter of $f(\vx)$ and $L$ is
  the Lipschitz constant of $\nabla f(\vx)$.
  \begin{proof}
    Dropping the subscript $m$ from $(\vx_m, t_m, \vd_m)$, we first observe that
\[
f(\vx+t\vd) \le f(\vx) + \langle \nabla f(\vx), \vd \rangle t + \frac{L}{2}\|\vd\|^2 t^2,
\]
by Lipschitz continuity.  Substituting $\vd =- \mf{H}^{-1}  \nabla  f(\vx)$ gives
\[
f(\vx+t\vd) = f(\vx) - \nabla f(\vx)^T  \mf{H}^{-1}  \nabla  f(\vx) t + \frac{L }{2}\|\mf{H}^{-1}\nabla f(\vx) \|^2 t^2.
\]
Letting $\lambda_{max}$ and $\lambda_{min}$ denote the
maximum and minimum eigenvalues of $\mf{H}$ evaluated at $\vx$, it also follows that
\[
  f(\vx+t\vd) \le f(\vx) - \frac{1}{\lambda_{\max}} \|\nabla f(\vx)\|^2 t  + 
    \frac{L }{2}  \frac{1}{\lambda^{2}_{\min}} \|\nabla f(\vx)\|^2 t^2.
\]
Letting $\bar t$ denote the  minimizer of the right-hand-side, we
conclude that
    \[
      f(\vx+t\vd) \le  f(\vx + \bar t\vd ) \le f(\vx) - \frac{1}{2}(\frac{\lambda^{2}_{\min}}{\lambda^2_{\max} L} 
      \|\nabla f(\vx)\|^2),
    \]
where the first inequality follows from the exact line search
used to select $t$.
Since  $\sigma^2 \ge \lambda^{2}_{\max}/\lambda^2_{\min}$,
we conclude that
\[
   f(\vx + t \vd) \le f(\vx) - \frac{1}{2\sigma^2 L}\| \nabla f(\vx) \|^2.
\]
    On the other hand, letting $f_* = f(\vx_*)$ 
    we have from strong convexity that the Polyak-Lojasiewicz
    inequality holds:
    \[
  \|\nabla f(\vx)\|^2 \ge 2\mu (f(\vx)-f_* ).
    \]
Hence,
\[
  f(\vx + t \vd)   \le f(\vx) - \frac{\mu}{\sigma^2 L} (f(\vx)-f_* ).
\]
Subtracting $f_*$ from both sides and factoring shows
\[
  f(\vx + t \vd)  - f_* \le  (1-\frac{\mu}{\sigma^2 L})(f(\vx) - f_*).
\]
It follows that each iteration $m$ satisfies
\[
  f(\vx_{m}) - f_* \le  (1-\frac{\mu}{\sigma^2 L})^{m}(f(\vx_{0}) - f_*).
\]
Since $\sigma \ge 1$ and $L \ge \mu$, the iterations converge, and the proof is completed.
  \end{proof}
\end{lemma}

Combining these lemmas shows that SAP globally convergences at (at least) a
linear rate.  By observing that SAP reduces to Newton's
method when the gradient is differentiable, we can also prove local quadratic
convergence assuming differentiability on a neighborhood of
the optimum $\vx_*$. 
\begin{theorem}
  The following statements hold.
  \begin{itemize}
    \item SAP globally converges from all initial conditions.
    \item If $\nabla f(\vx)$  is differentiable on the ball $B(\vx_*, r) := \{ \vx :  \|\vx-\vx_*\| \le r\}$
  for some $r > 0$, then SAP exhibits quadratic convergence, i.e., for some finite
  $M$ and $\zeta > 0$
  \[
    \|\vx_{m+1} - \vx_*\| \le \zeta \|\vx_{m} - \vx_*\|^2
  \]
  for all $m > M$.
  \end{itemize}

  \begin{proof}

    The first statement follows from Lemmas~\ref{lem:PropertiesOfObj}~and~\ref{lem:GlobalConv}.

  To prove the second, we show that $B(\vx_*, r)$
    contains a sublevel set $\Omega_{\beta} = \{ \vx : f(\vx) \le \beta\}$ for some
  $\beta > 0$,  implying that SAP 
  reduces to Newton's method with exact line search for some $m > M$,
 given that sublevel sets are invariant. 

  To begin, we have, by strong convexity, that
  \begin{equation}
    \beta \ge f(\vx) \ge f(\vx_*)  + \frac{\mu}{2} \|\vx-\vx_*\|^2,
    \label{eq:strong_convexity_at_differentiable_optimal}
  \end{equation}
    for all $\vx \in \Omega_{\beta}$.
    Rearranging shows that
    \[
      \|\vx-\vx_*\|^2 \le 2\frac{\beta- f(\vx_*)}{\mu}.
    \]
    Hence, $B(\vx_*, r)$ contains
    $\Omega_{\beta}$ for any $\beta$ satisfying $2\frac{\beta- f(\vx_*)}{\mu} < r$.
    For some finite $M$, we also have that $v_m \in \Omega_{\beta}$ for all $m > M$ 
    by Lemma~\ref{lem:GlobalConv}.

    Next, we prove that Newton iterations are quadratically convergent
  with exact line search. Indeed, using once more the strong convexity result in Eq.~\eqref{eq:strong_convexity_at_differentiable_optimal}
  \begin{align*}
    \|\vx_{m+1} - \vx_*\|^2&\le \frac{2}{\mu} ( f(\vx_{m+1}) - f(\vx_*)) \\
                    &= \frac{2}{\mu} ( f(\vx_m + t_m \vd_m   ) - f(\vx_*)) \\
                    &\le \frac{2}{\mu} ( f(\vx_m + \vd_m   ) - f(\vx_*)) \\
                    &\le \frac{2}{\mu} L \|\vx_m + \vd_m - \vx_*\|^2,
  \end{align*}
    where the first line uses strong convexity,
    the third exact line search, and the last
    Lipschitz continuity. But for some $\kappa > 0$,
    we have that $\|\vx_m + \vd_m - \vx_*\|^2 \le \kappa \|\vx_{m} - \vx_*\|^4$
    by quadratic convergence of Newton's method with unit step-size (\cite[Theorem 3.5]{bib:nocedal2006numerical}).
    Hence,
    \[
      \|\vx_{m+1} - \vx_*\|^2 \le \frac{2}{\mu}  L \kappa \|\vx_{m} - \vx_*\|^4,
    \]
    and the claim is proven.
  \end{proof}
\end{theorem}

% Acknowledgment: N.B. Use section*.
\section*{Acknowledgment}
The authors would like to thank especially to Michael Sherman for his trust on
this research from day one and to the Dynamics \& Simulation and Dexterous
Manipulation teams at TRI for their continuous patience and support.

% Can use something like this to put references on a page by themselves when
% using endfloat and the captionsoff option.
\ifCLASSOPTIONcaptionsoff
  \newpage
\fi

% References:
\bibliographystyle{./IEEEtran/IEEEtran}
\bibliography{sap_paper}

\end{document}